\documentclass[11pt]{article}
\usepackage[utf8]{inputenc}

\usepackage{hyperref}
\usepackage{enumitem}
\usepackage{url}
\usepackage{times}
\usepackage{framed}
\usepackage[british]{babel}
\usepackage{subcaption}
\usepackage{commath}
\usepackage{siunitx}
\usepackage{amsmath}

\usepackage[ruled,vlined]{algorithm2e}
\usepackage{graphicx}

\SetCommentSty{mycommfont}

\usepackage[apaciteclassic]{apacite}

\usepackage{float}

\title{Using Score Distributions to Compare Statistical Significance Tests for Information Retrieval Evaluation}

\begin{document}

\author{{\small \begin{tabular}{cccc}
  Javier Parapar$^*$ & David E. Losada$^\dag$ & Manuel A. Presedo-Quindimil$^*$ & Alvaro Barreiro$^*$ \\
	&&&\\
   \multicolumn{2}{c}{$^*$IRLab, Dept of Computer Science} &      \multicolumn{2}{c}{$^\dag$Centro Singular de Investigación}  \\
  \multicolumn{2}{c}{$^\dag$Universidade da Coru\~na} &   \multicolumn{2}{c}{en Tecnolox\'ias da Informaci\'on (CiTIUS)}  \\
    \multicolumn{2}{c}{Campus de Elvi\~na} &    \multicolumn{2}{c}{Universidade de Santiago de Compostela}     \\
     \multicolumn{2}{c}{15071, A Coru\~na (Spain)} &     \multicolumn{2}{c}{Campus Vida}   \\
 \multicolumn{2}{c}{} &    \multicolumn{2}{c}{15782, Santiago de Compostela (Spain)}  \\
	&&&\\
 javierparapar@udc.es	& david.losada@usc.es & mpresedo@udc.es & barreiro@udc.es
	 \end{tabular}}}

\date{\small{(Submitted to JASIST Oct 11, 2017; 1st revision May 22, 2018; 2nd revision Sep 5, 2018; accepted Jan 11, 2019)}}

\maketitle

\begin{abstract}
Statistical significance tests can provide evidence that the observed difference in performance between two methods is not due to chance. 
In Information Retrieval, some studies have examined the validity and suitability of such tests for comparing search systems.
We argue here that current methods for assessing the reliability of statistical tests suffer from some methodological weaknesses,
and we propose a novel way to study significance tests for retrieval evaluation. Using Score Distributions, we model the output of multiple search systems, produce simulated search results from such models, and compare them using various significance tests. 
A key strength of this approach is that we assess statistical tests under perfect knowledge about the truth or falseness of the null hypothesis. 
This new method for studying the power of significance tests in Information Retrieval evaluation is formal and innovative. 
Following this type of analysis, we found that both the sign test and Wilcoxon signed test have more power than the permutation test and the t-test.
The sign test and Wilcoxon signed test also have a good behavior in terms of type I errors. The bootstrap test shows few type I errors, but it has less power than
the other methods tested. 
\end{abstract}

\section{Introduction}

\label{sec:intro}

Use of significance tests is a de facto standard of evaluation in Information Retrieval (IR). 
In the early days of IR experimentation, researchers tended to prefer the 
Wilcoxon signed-rank test and the sign test, which are simple
and make few assumptions about the data. Other parametric alternatives, such
as Student's t-test, require data drawn from specific distributions (e.g. Gaussian), but the output of retrieval experiments violates this assumption. 

A number of studies have analysed the reliability of significance tests \cite{Zobel:1998,Voorhees2002,Urbano:2013,Sanderson:2005,Cormack2007,Sakai:2016:TST:2911451.2914684,Smucker:2007}. 
Many of them suggest that, despite its assumptions, the t-test performs well and
it should be preferred over the Wilcoxon signed-rank test and the sign test. 
Smucker and colleagues argued that the use of the 
Wilcoxon signed-rank test and the sign test should cease \cite{Smucker:2007}, and 
have influenced a change towards the t-test and the permutation test.
The way in which experimenters employ statistical testing is crucial and
affects the dissemination of research results, as it has been demonstrated in other fields \cite{Miettunen2003}. 
A comprehensive and solid analysis of the relative merits of each significance test is therefore
essential for IR experimentation.

A significance test stands on a null hypothesis ($H0$) and an alternative hypothesis ($H1$). In IR, many experiments run
paired two-sided tests. In such a case, the null hypothesis states that the two retrieval outputs under examination are drawn from the same
population (any difference between them is due to chance), and 
the alternative hypothesis states that they are drawn from different populations. The test estimates the probability $p$ (p-value) of observing
a difference at least as large as that produced by the experiment given that $H0$ is true. 

The studies on the reliability of significance tests for IR evaluation have followed two main methodologies. {\em Query splitting}
methods split the topics of a test collection into two groups, run each significance test in each group, and check the coherence
of the results. Smucker and colleagues followed an alternative approach \cite{Smucker:2007}, where the p-values estimated by 
a permutation (or randomisation) test are the main reference and other significance tests are evaluated  {\em in comparison to the permutation test}. We claim here that both methodologies have severe limitations and propose an innovative way
for assessing significance tests. 

Query splitting methods are limited to assess the consistency of a significance test with itself. But the results of the test
can be consistently wrong over the splits (rejecting a true $H0$ --type I error-- or failing to reject a false $H0$ --type II error--). 
With no knowledge of the truth or falseness of $H0$, we should not just equate consistency with reliability. 
Comparing significance tests based on the permutation test, as done in \cite{Smucker:2007}, is not exempt from problems either. The permutation test
can compute a good approximation to the exact p-values, but we should not produce miss or false alarm rates from such p-values. 
Given a certain significance level ($\alpha$) and the p-values estimated by permutation, the miss and false alarm
rates of each significance test are measured based on the agreement between the test's decisions and the permutation test's decisions. 
Such an approach implicitly assumes that those cases where the p-value produced by permutation is above $\alpha$
are cases where $H0$ is true and, conversely, those cases below $\alpha$ are cases where $H0$ is false. 
This rule compromises the quality of the analysis because the permutation test is not error-free. 
For example, with $\alpha=.05$, the permutation test would be making an average
of $5\%$ type I errors (no difference between the systems, but the permutation test says otherwise). 
As such, in 5\% of cases, giving blind faith to the permutation test unfairly penalises any significance test that 
makes the correct decision (any significance test that is skeptical about the difference is actually right!). 
Likewise, the permutation test makes some type II errors and accepting such permutation test's decisions is unfair to those tests that detect the difference. 
In summary, the main flaw of existing methods to assess the reliability of significance tests is that they make strong assumptions about the truth value of the
null hypothesis. Although the previous studies substantially contributed to analysing the use of significance tests in IR, we believe that a more robust methodology,
based on actual knowledge about $H0$, can be designed. This is precisely the primary aim of our paper.

%


We propose a method for assessing the reliability of significance tests that works with simulated results of retrieval systems. We take many executions (runs) from TREC systems, and we model IR systems using Score Distributions (SDs) \cite{Manmatha2001,Arampatzis:2011} learnt from those real runs. 
With the modeled systems, we can produce multiple retrieval results by sampling from the distributions. 
The core idea is that we can compare significance tests under complete knowledge
about the null hypothesis. For example, we can experiment with the same system producing two ranked lists (null hypothesis true) 
or we can experiment with two different systems producing each a ranked list (null hypothesis false). Systems with distinctive retrieval characteristics
can be obtained by manipulating the parameters of the inferred statistical distributions. Furthermore, this simulation method can be repeated as many times as needed, serving to reinforce the validity of the study. 

We examined the tests with this innovative methodology and found some results that 
agree with those presented by Smucker and colleagues \cite{Smucker:2007}: the permutation and the t-test tend to agree with each other and, to a lesser degree, with bootstrap; while
the Wilcoxon test and the sign test disagree with both the permutation and the t-test. 
However, we found substantive evidence that suggests that these differences occur because the Wilcoxon and the sign test have higher power than the permutation test. 
These results are in agreement with the findings reported by authoritative statistical studies, which showed the relative lack of power of the permutation test (see Section 5.11 \cite{conover99}). 


The main contributions of this paper are:

\begin{enumerate}
	\item  a new method for assessing the reliability of the statistical tests based on the analysis of their power using data derived from simulated IR systems.
	\item a complete empirical study that categorically concludes that, in typical IR experimentation, the Wilcoxon and sign tests are more powerful than the permutation, bootstrap or the t-test. 
\end{enumerate}


\section{Related Work}

Non-parametric tests of significance, such as the Wilcoxon test and the sign test, have been widely used in IR experiments. Both tests assume data drawn from continuous distributions, while 
retrieval experiments produce discrete data. Despite this fact, Van Rijsbergen  suggested conservative use of such tests \cite{Rijsbergen:1979}.
The t-test has been also used regularly in IR. Hull \cite{Hull:1993} claimed that it often performs well even when the normality assumption is violated. Other popular tests 
in IR are the bootstrap test \cite{Savoy1997495,Cormack:2006} and the permutation test \cite{Smucker:2007}.

Several papers studied the reliability of the significance tests. Zobel \cite{Zobel:1998} made multiple pairwise-comparisons of systems with two disjoint query sets.
A type I error was recorded when a test observed a significant difference between two systems on the first query set and the ordering of the systems was different on the second query set. 
He concluded that the Wilcoxon test is more reliable and has higher power than both the t-test and ANOVA. 
Sanderson and Zobel \cite{Sanderson:2005}, who expanded a previous study by \cite{Voorhees2002}, found that the t-test shows lower error rates when compared to the sign and the Wilcoxon tests. 
Cormack and Lynam \cite{Cormack2007} also performed a query-splitting study and observed that the t-test, Wilcoxon and the sign test are highly accurate and have high power, with the t-test proving superior overall. 
Urbano et al. \cite{Urbano:2013} also presented a comparison of significance tests based on query-splitting. Regarding {\em safety} (smallest error rates across significance levels), 
the t-test was the best, followed by the Wilcoxon test (for low levels of significance) and the permutation test (for the usual levels of significance). 
The bootstrap test consistently produced smaller p-values and, thus, it was the most powerful across significance levels. 
The authors also studied the agreement of the tests with themselves, and the Wilcoxon test turned out to be the most stable of all for small p-values, while the t-test the most stable overall. 
The permutation test was also evaluated in this study, but it was not optimal under any criterion considered. 
All these studies provide valuable evidence for IR experimentation. However, the query-splitting methodology makes an arbitrary division of topics and lacks real knowledge about the truth
or falseness of the null hypothesis. The fact that a significance test gives consistent results over two splits of topics should not be considered a measure of quality. 
As a matter of fact, the test might be consistently wrong over the splits (consistently rejecting a true null hypothesis or consistently accepting a false null hypothesis).


Smucker and colleagues \cite{Smucker:2007} found that the bootstrap test, the t-test and the permutation test produce comparable p-values, and showed that both the Wilcoxon and the sign test
are discordant with the permutation test. Following their experiments, they suggested discontinuing the use of these two tests. 
Our study also reveals a discordance between these two tests and the rest of the tests, but we argue that the output of the permutation test must not be taken as the main reference. Smucker et al.'s study 
evaluated significance tests in terms of how well each test matched the decisions of the permutation test. For example, a given test was assigned a false alarm when the test labeled a difference as significant while the permutation test considered it as non-significant (see e.g. Table 2, \cite{Smucker:2007}). We claim that assigning this kind of {\em ground truth role} to the permutation test limits the conclusions that can
be drawn from the analysis. 
We show here that the Wilcoxon test and the sign test have higher power than the other tests and make a low number of type I errors. 
Therefore, significant differences found with these two tests must not be ignored on the basis that other tests fail to identify these differences.

Sakai \cite{Sakai:2016:TST:2911451.2914684} compared two versions of two-sample t-tests: Student's t-test 
and Welch's t-test. He also employed a query-splitting methodology but his approach has some restricted knowledge about the null hypothesis. 
More specifically, he modeled the case of $H0$ being true as follows. Given a query set of size $n$ and $m$ runs that processed these queries,
the queries are randomly partitioned into two sets. For each of the runs and a given evaluation metric, he conducted a two-sided, two-sample test to determine whether or not 
the difference between the two means for the same run are statistically significant. The ground truth was that they are not, since the scores actually come from the same system. This strategy, based on unpaired data, cannot be applied to evaluate significance tests under
current TREC-like exercises (where all systems are evaluated with the same queries). However, as argued by Sakai, possible applications
of two-sample tests in IR include comparing set of clicks from two search engines or comparing the difficulty of two test collections using
the same search system.


Our paper is also related to some studies that also employed simulation in order to analyze other aspects of IR evaluation. 
For example, Urbano \cite{Urbano2016} performed a study on test collection reliability that compared a number of measures and estimators of test collection accuracy. His method was 
based on stochastic simulation of evaluation results and, through large-scale simulation from TREC data, the bias of several estimators of test collection accuracy was analyzed.

Score distributions \cite{Arampatzis2009} have been studied and modeled since the early days of IR \cite{J.A.Sweets1963}. 
Different combinations of statistical distributions have been proposed for modeling the score distributions
of relevant and non-relevant documents (two Gaussians \cite{J.A.Sweets1963}, two negative exponentials \cite{Swets1969}, two Poissons \cite{Bookstein1977}, two Gammas \cite{Baumgarten:1999}, a Gaussian and a negative exponential \cite{Arampatzis2000,Manmatha2001,a2009}, or a Gaussian and a Gamma \cite{Kanoulas2009,dai2011variational}). Some studies \cite{Kanoulas2010} have also analysed
SDs based on the scoring formulas of the retrieval models. 
Manmatha et al. \cite{Manmatha2001} proposed the use of SDs to combine the outputs of multiple search engines. Arampatzis et al. \cite{Arampatzis2000,a2009} 
utilised SD models for threshold optimisation in a legal search task. Cummins employed SDs for query performance prediction \cite{Cummins:2014:DSD:2576772.2559170}. Arampatzis et al. experimented with SDs in image retrieval \cite{ipm1}, Parapar et al. employed SDs in pseudo-relevance feedback \cite{Parapar2014}, and Losada at al. proposed a rank fusion approach based on SDs for prioritising assessments in IR evaluation \cite{losadaetal17b}.

\section{Analysing Significance Tests with Score Distribution Models} 
\label{astsdm}


Score distributions model the way in which search systems generate retrieval scores. We can, therefore, simulate multiple search systems and evaluate their output using 
Average Precision. Different conditions --null hypothesis true/false-- can be generated, and significance tests can be evaluated accordingly. Our method proceeds as follows:

\begin{enumerate}
	
	\item For every (TREC run, query) pair, we learn a SD model (a mixture of statistical distributions) from the list of scores supplied by the run. 
	\item For each run we take its 50 (as many as different queries) mixtures and we experiment with the case of a true null hypothesis (same model producing two outputs): 
	\begin{enumerate}
		\item we randomly extract 1000 samples from the SD model and obtain a \textit{synthetic} list of scores and their relevance values 
		(the method extracts samples from either the distribution of relevant documents or the distribution of non-relevant documents and, thus, 
		each extracted score is assigned a relevance label, 1 if the score came from the relevant document distribution and 0 if the score came from the non-relevant document distribution).
		The resulting list of scores is sorted in descending order. 
		
		\item we repeat step (a) and obtain a second \textit{synthetic} list.
		\item we compute the average precision of both lists.
	\end{enumerate}
		\item[] Given the two sequences of 50 APs obtained, a significance test is run for assessing the significance of the difference found.  
	
	\item Step 2 is repeated 1000 times and we record the average number of times that the significance test falsely rejects $H0$.
	
	\item Now, we experiment with the case of a true alternative hypothesis (the outputs come from different models). For each run we take its 50 (as many as different queries) mixtures and
	\begin{enumerate}
		\item we randomly extract 1000 samples from the SD model and obtain a \textit{synthetic} list of scores and their relevance values.
		The resulting list of scores is sorted in descending order. 
		
		\item we alter the SD model's parameters, sample from the modified model and obtain a second \textit{synthetic} list.
		\item we compute the average precision of both lists.
	\end{enumerate} 
	\item[] Given the two sequences of 50 APs obtained, a significance test is run for assessing the significance of the difference found.  
	 \item Step 4 is repeated 1000 times and we record the average number of times that the significance tests correctly rejects $H0$.
	 \item Steps 4-5 are repeated several times by gradually separating the modified model from the original model (parameter manipulation of the mixture). 	
\end{enumerate}

With this procedure, we can straightforwardly estimate the power of a significance test, $p(Reject$ $H0|H0$ $is$ $false)$, 
the probability of a type I error, $p(Reject$ $H0|H0$ $is$ $true)$, and so forth. 
The pseudo-code that implements this procedure is available in the Appendix.


\subsection{Modeling Information Retrieval Systems}

With real search systems, a comparative analysis of significance tests is tricky. We have no knowledge of 
the underlying retrieval models that generate the search results and, thus, we know nothing about the truth of the null hypothesis. 
Furthermore, we often observe a small number of executions from each search system. Such limited sample imposes limitations on the statistical analysis. 
By modeling search systems with statistical models we can produce as many samples as required and, additionally, we have certainty about the null hypothesis.

%

Score distribution models assume that the distribution generating the scores of relevant documents is different from the one producing the scores of non-relevant documents. 
Various combinations of distributions have been employed to model each group of documents, and 
the parameters of the mixture distribution can be learned from the observed documents' scores. 
We employ here a two log-normal distribution \cite{Cummins2011}, which adheres to the recall-fallout convexity hypothesis \cite{Robertson2007}, and shows higher goodness of fit when compared with other alternatives \cite{Cummins2011,Cummins:2012:TVS:2260641.2260691}.


Each retrieval system has a set of mixtures (one mixture per query). For every query $q^{(i)}$, the mixture has two log-normal distributions: $L_0^{(i)}$ for the non-relevant documents and $L_1^{(i)}$ 
for the relevant documents. If $P(s|1)^{(i)}$ is the probability density function (pdf) for the scores ($s$) of relevant documents and $P(s|0)^{(i)}$ is the pdf for the 
scores of non-relevant documents then the mixture is: 

\begin{equation}
P(s)^{(i)} = \lambda^{(i)} P(s|1)^{(i)} + (1-\lambda^{(i)}) P(s|0)^{(i)}
\end{equation}

\noindent where $\lambda^{(i)}$, the mixture parameter, represents the proportion of relevant documents returned by the system for query $q^{(i)}$.

Some TREC runs produce negative scores and we followed the standard procudure of shifting all scores of these runs by some constant factor.
We discarded those runs that do not supply retrieval scores for the returned documents. 
The use of alternative SD models that do not rely on the existence of scores \cite{Robertson:2013:MSD:2499178.2499181} is left to future work. 

For each query $q^{(i)}$, TREC supplies a set of relevance assessments. Given the TREC run and the relevance assessments, we proceeded to learn the run's SD model
as follows. The mixing weight is set to the proportion of relevant documents returned by the run.  
The scores of the relevant documents returned by the run are used to learn the parameters of $L_1^{(i)}$. This learning stage was done following
a maximum likelihood approach\footnote{We performed direct optimization of the log-likelihood with the Nelder-Mead method  \cite{Nelder1965} using the \texttt{fitdistrplus} R package: http://cran.r-project.org/web/packages/fitdistrplus/fitdistrplus.pdf.} \cite{Dempster77maximumlikelihood}, which 
provides better goodness of fit than the method of moments \cite{Cummins2011}.
Following usual practice in pooling-based IR, the set of non-relevant documents retrieved by the run is composed of the judged non-relevant documents plus the unjudged documents.
The procedure to learn $L_0^{(i)}$ from this set was the same than that used to learn $L_1^{(i)}$ from the set of relevant documents.

%



The ultimate objective of our study is to analyse significance tests with a series of per-query results. The procedure sketched above leads to SD models that can generate multiple system outputs in the form of lists of retrieval scores. Classic IR measures, such as Mean Average Precision ($MAP$), rely on relevance judgments at document level. 
By sampling separately from the two component distributions (e.g., see step 2(a) in the procedure sketched above), we ensure that each sampled score is assigned a relevance value. In this way, the resulting
ranking of scores can be evaluated with any performance measure, such as MAP.

\subsection{Power Curves}

\begin{figure}
	\centering
	\includegraphics[width=0.7\columnwidth]{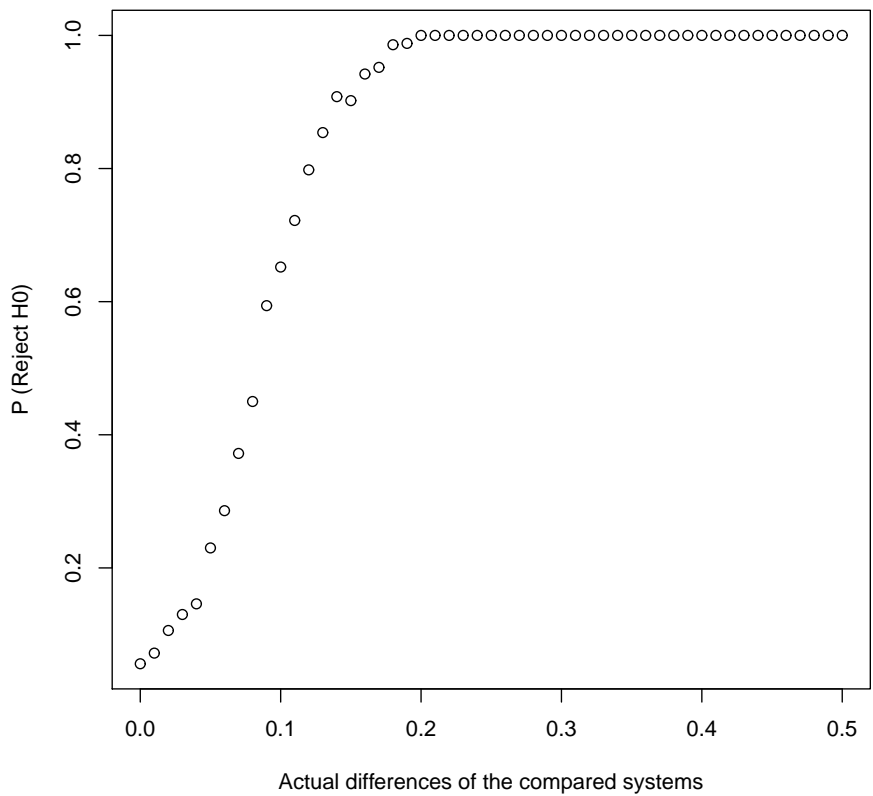}
	\caption{Example of Power Curve for the Wilcoxon Test. The X axis represents increasing differences between the systems being compared. The null hypothesis, H0, only holds in the leftmost point, the other points correspond with H0 false.  }
	\label{power}
\end{figure}

Given a certain level of significance, a common way to analyse significance tests consists of studying the form of the power curve. 
A power curve plots the probability of rejecting $H0$ against increasing differences between the systems being compared. 
Figure \ref{power} shows an example of a power curve. The y-axis represents the probability of rejecting the null hypothesis ($P(Reject\ H0)$), 
while the x-axis represents the difference between the systems that are compared. 
The $0.0$ point of the x-axis (leftmost point) corresponds to the case where $H0$ is true ($x=0.0$, no difference between the systems). 
The rest of the points of the x-axis ($x>0.0$) correspond to cases where $H0$ is false (and the larger $x$ is, the more different the systems are). 
The height of the leftmost point
represents the probability of a type I error (incorrect rejection of the null hypothesis). Typically, the power curve starts 
from a height equal to the significance level, rises smoothly and monotonically and, eventually, reaches the maximum probability of 1. 
The closer the curve is to a right angle, the more power the test has.

%

Such an analysis must only be done under certainty about the hypothesis being tested. Previous studies on significance tests for IR lack such certainty. By design, our methodology produces
outputs where we know whether or not the test must reject $H0$. To simulate the case when $H0$ is true 
we just have to obtain two random samples from the same SD model (same system producing two outputs), 
and compare the two series of APs (one pair of samples per query). 
To simulate the case when $H0$ is false, we proceed by taking a SD model and altering its parameters. In this way, we can compare the original SD model against a modified SD model. Given the outputs
produced by these two models, we are certain that the series of APs come from different statistical distributions. 
Note that the modifications are done on a per-query basis because each system has a SD model learnt for each query. 
There are multiple possibilities to modify the original SD model. For example, by changing $\lambda^{(i)}$ we can simulate 
the increase or decrease in the number of relevant documents returned. We decided to leave the same $\lambda^{(i)}$ for both models because
changing $\lambda^{(i)}$ models a change in the proportion of relevant documents returned. In TREC, the number of relevant documents
is fixed --for each query, we only have assessments for the pooled documents-- and, as a consequence, it does not make sense to model a system that
returns more relevant documents than those available in the pool. Another possibility consists of changing the mean of $L_{0}$ (or $L_{1}$), which has the effect of altering the position in the ranking of the non-relevant (or relevant) documents. We opted to gradually increase the mean (location) of $L_{1}$ ($\mu_1$). This choice simulates relevant documents moving up in the ranking and, thus, improves the effectiveness of the original model unequivocally and monotonically. Our report of significance tests will, therefore, include
power curve plots where the x-axis represents the percentage of increase in $\mu_1$.
In Section \ref{s:validity}, we provide empirical evidence showing that increasing $\mu_{1}$ leads to improvements in AP.


\section{Experiments}
\label{s:exp}

We performed a thorough analysis of several significance tests using the ad-hoc retrieval runs submitted to TREC 3, 5, 6, 7 and 8 \cite{Voorhees:2005:TEE:1121636}. 
The number of systems that we used in our experimentation (see Table \ref{t_trec}) is lower than the official number of participants
because we removed the systems that did not return retrieval scores or retrieved very few documents per query. 
We worked with the top 1000 documents ranked by the systems and we set the number of random samples generated from the SD models to 1000.

\begin{table}
	\centering
	\caption{Summary of TREC system runs used in the experiments.}{
		\begin{tabular}{l|c|c|c|c}
			Edition & Query set & Relevant Documents & Systems & Used systems \\\hline
			TREC 3 & 151-200 & 9805 & 40 & 32\\
			TREC 5 & 251-300 & 5524 & 61& 38\\
			TREC 6 & 301-350 & 4611 &74& 48\\
			TREC 7 & 351-400 & 4674 &103& 62\\
			TREC 8 & 401-450 & 4728 &130& 88\\ \hline
		\end{tabular}
	}
	\label{t_trec}
\end{table}

\subsection{Significance Tests}
\label{s:test}

We compared the following tests: t-test, Wilcoxon signed rank test, sign test, permutation test and bootstrap. The comparison was done for the two-sided case using paired data. 

The t-test assumes that the obtained differences follow a normal distribution. The null hypothesis is that the mean of the distribution of differences is zero. Wilcoxon assumes that the differences can be ranked, but ignores the magnitude of the differences. The rank values are assigned the sign of the measured difference, and the null hypothesis is that the sum of the positive ranks is the same as the sum of negative ranks. The sign test relies on even less stringent assumptions: under the null hypothesis, we would expect the same number of positive and negative differences. The permutation test is free of mathematical assumptions. The null hypothesis is that the two systems are identical and any permutation of the matched pair observations will produce an output equally probable. Given a statistic for the test and computing all possible permutations of the observed values it is possible to compute the exact p-value. The null hypothesis of the bootstrap test is that the observed values are random samples from the same distribution. 


For the t-test, Wilcoxon, sign test, and permutation test we used the implementation of the JSC (Java Statistical Classes) by Andrew Bertie from The Open University. 
For bootstrap, we implemented the test as described in \cite{Efron1993} ({\em the one-sample problem}). Following \cite{Smucker:2007}, we took 
the difference of means as the statistic for both permutation and bootstrap, and we extracted 100.000 samples of random permutations. We also considered the popular t-test statistic
for permutation and bootstrap but these two tests performed better when the difference of means was the statistic.


\subsection{Experiments: Type I Error and Power Curves}

\begin{figure*}
	\centering
	\begin{subfigure}[b]{0.49\textwidth}
		\includegraphics[width=\textwidth]{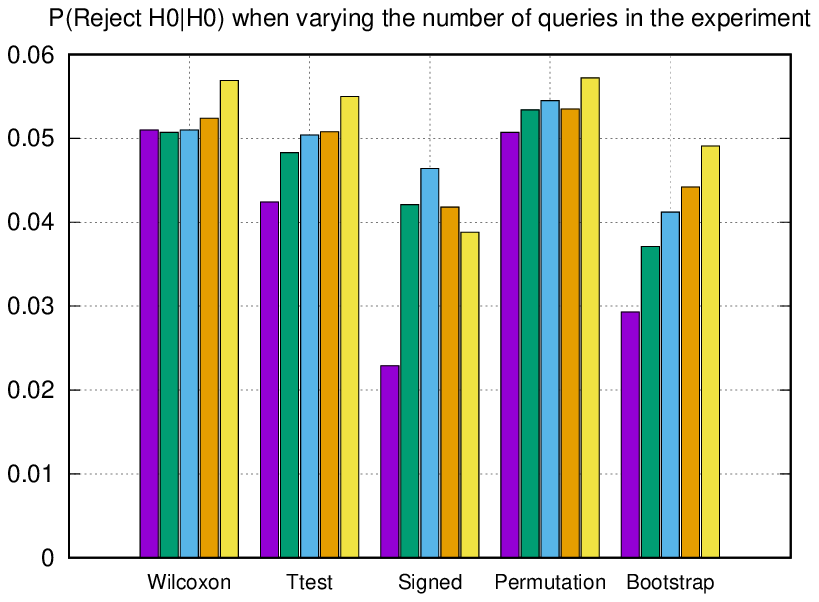}
		\subcaption{TREC 3}
		\label{t3:errortypeI}
	\end{subfigure}
	\begin{subfigure}[b]{0.49\textwidth}
		\includegraphics[width=\textwidth]{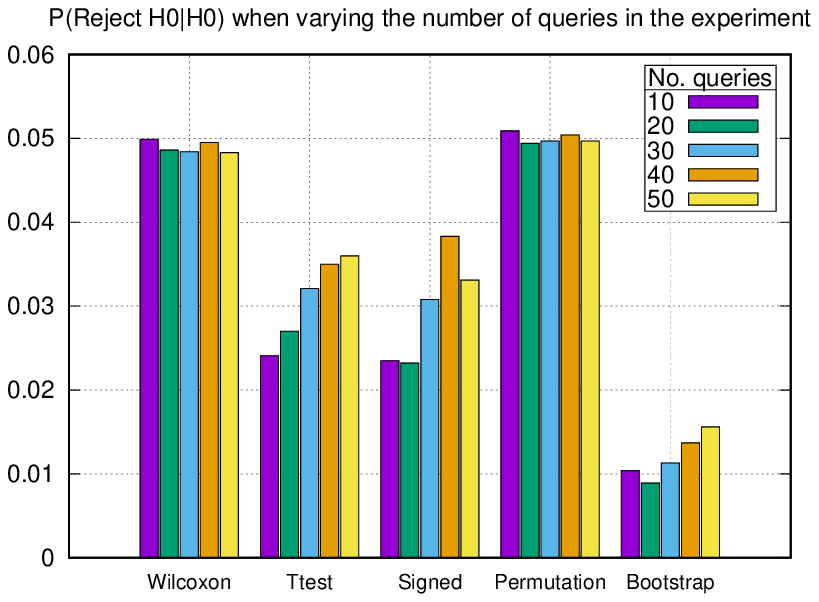}
		\subcaption{TREC 5}
		\label{t5:errortypeI}
	\end{subfigure}\\
	\begin{subfigure}[b]{0.49\textwidth}
		\includegraphics[width=\textwidth]{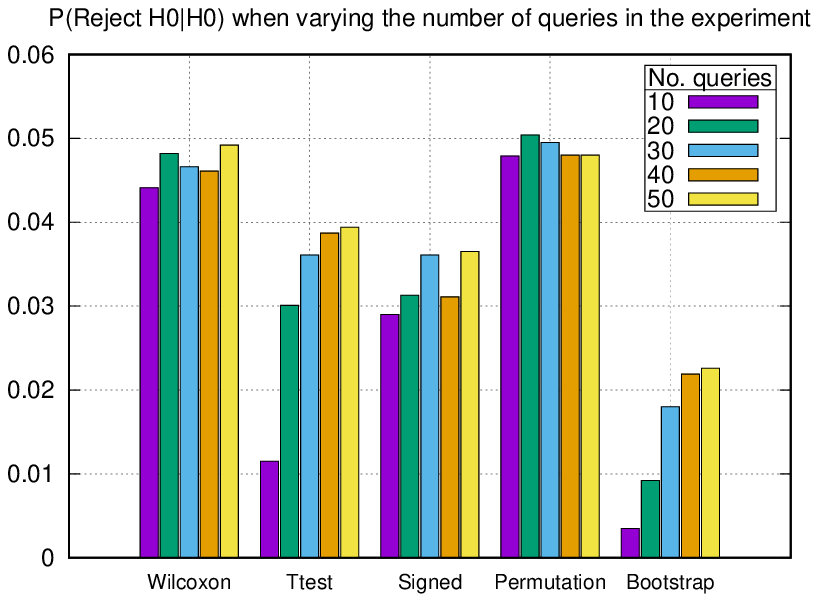}
		\subcaption{TREC 6}
		\label{t6:errortypeI}
	\end{subfigure}
	\begin{subfigure}[b]{0.49\textwidth}
		\includegraphics[width=\textwidth]{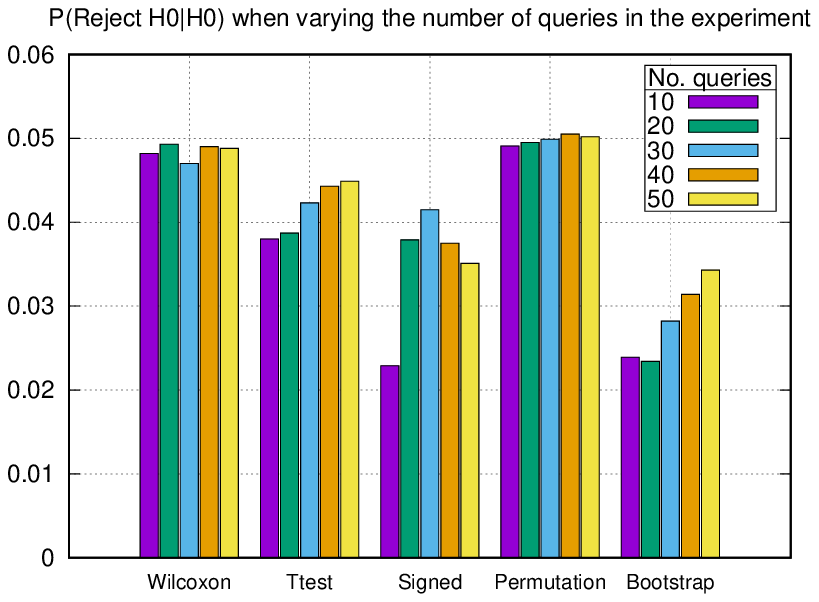}
		\subcaption{TREC 7}
		\label{t7:errortypeI}
	\end{subfigure}\\
	\begin{subfigure}[b]{0.49\textwidth}
		\includegraphics[width=\textwidth]{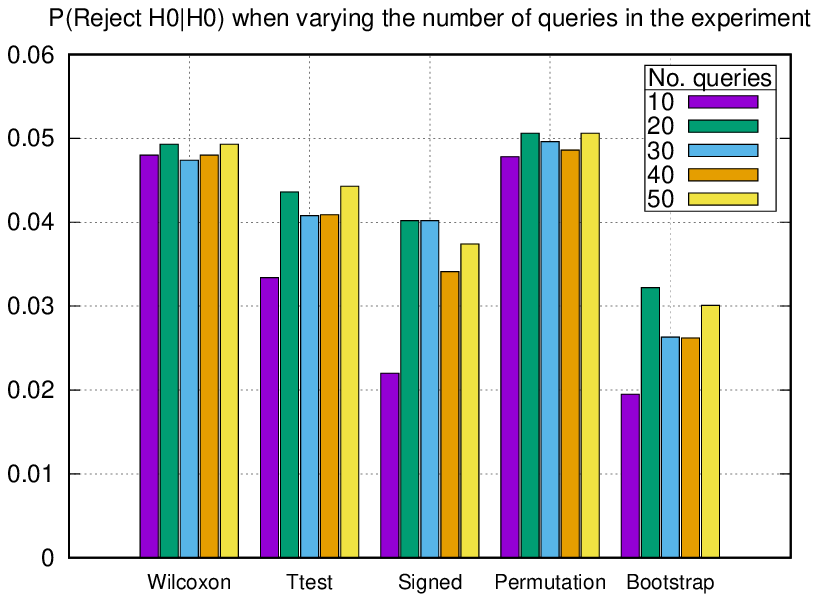}
		\subcaption{TREC 8}
		\label{t8:errortypeI}
	\end{subfigure}            
	
	\caption{Average $p(Reject$ $H0|H0$ $is$ $true)$ in different TREC collections ($\alpha=0.05$)}
	\label{typeIfigs}
\end{figure*}

To estimate the probability of a type I error, 
$p(Reject$ $H0|H0$ $is$ $true)$, we produced two samples from the same SD model and ran the significance test. 
This procedure was repeated 1000 times for each system and we averaged --over all TREC systems-- the number of times that $H0$ was rejected (the significance level was set to 0.05).
Figure \ref{typeIfigs} shows the results of this experiment with varying number of queries.

Wilcoxon and the permutation test tend to achieve the expected probability of rejecting $H0$ when $H0$ is true 
(the significance level). The other three tests are more conservative and show a probability of rejecting $H0$ lower
than the significance level. Such conservative behavior becomes apparent with small query sets. Bootstrap is extremely conservative (particularly in TREC 5 and 6, where
the retrieval performance of the systems is the lowest). By design, tests are expected to have $5\%$ of type I errors and, thus, the t-test, the sign test and bootstrap do not behave as expected. 
From a practical perspective, these three tests are making fewer errors, which might seem convenient. However, from a statistical standpoint, this result suggests that the p-values obtained by these tests are worse estimations of the probability of finding the observed difference when the null hypothesis is true.
We also experimented with other significance levels --from 0.01 to 0.25, steps of 0.01-- 
and found consistent results: Wilcoxon and the permutation test yielded results that matched with the significance level, while the other three tests made fewer errors than expected.

\begin{figure*}
	\centering
	\begin{subfigure}[b]{0.49\textwidth}
		\includegraphics[width=\textwidth]{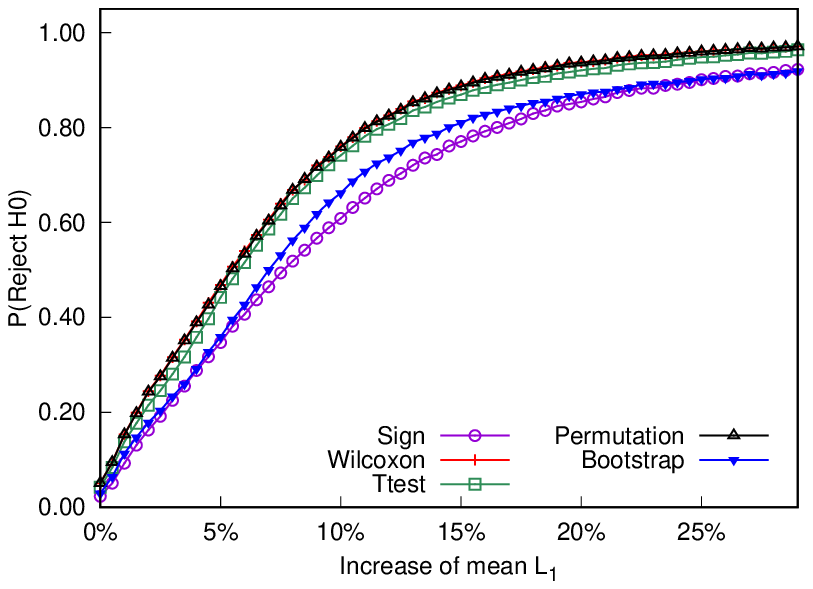}
		\subcaption{10 queries}
		\label{trec3q10}
	\end{subfigure}
	\begin{subfigure}[b]{0.49\textwidth}
		\includegraphics[width=\textwidth]{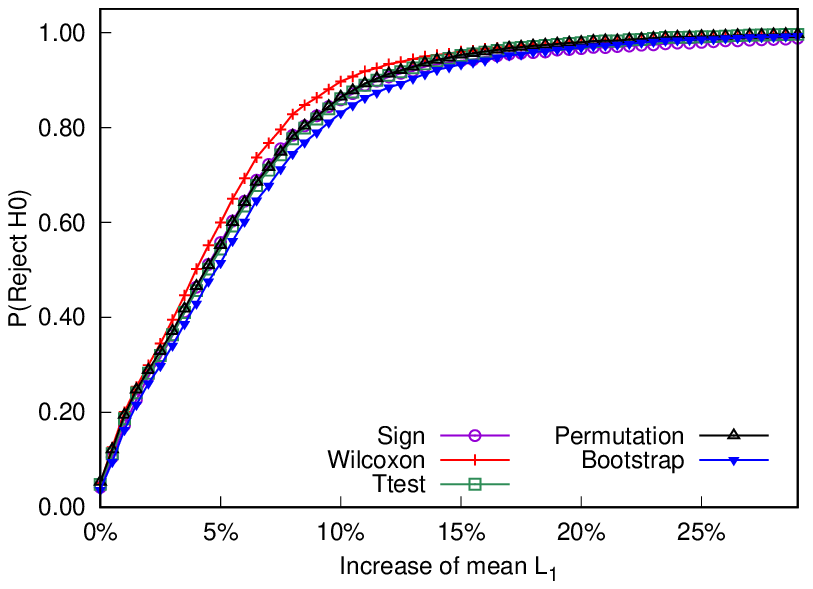}
		\subcaption{20 queries}
		\label{trec3q20}
	\end{subfigure}\\
	\begin{subfigure}[b]{0.49\textwidth}
		\includegraphics[width=\textwidth]{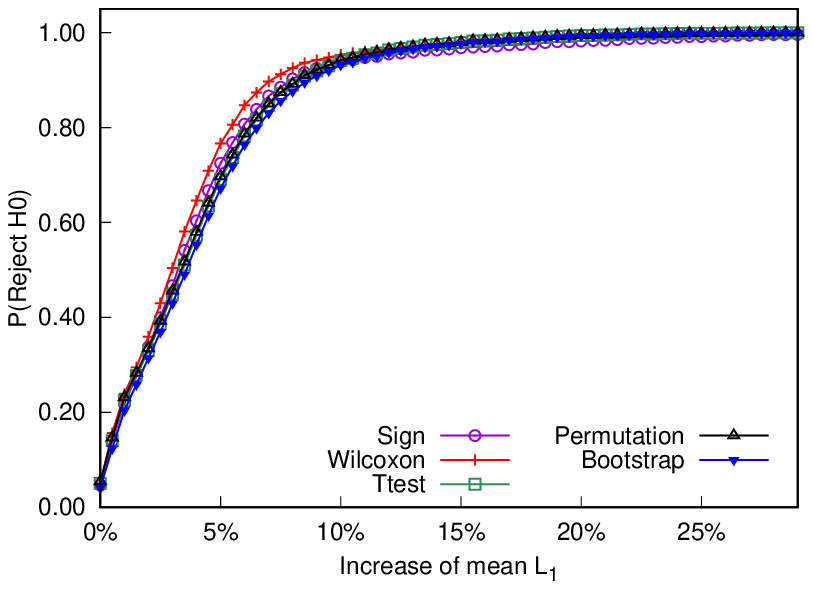}
		\subcaption{30 queries}
		\label{trec3q30}
	\end{subfigure}
	\begin{subfigure}[b]{0.49\textwidth}
		\includegraphics[width=\textwidth]{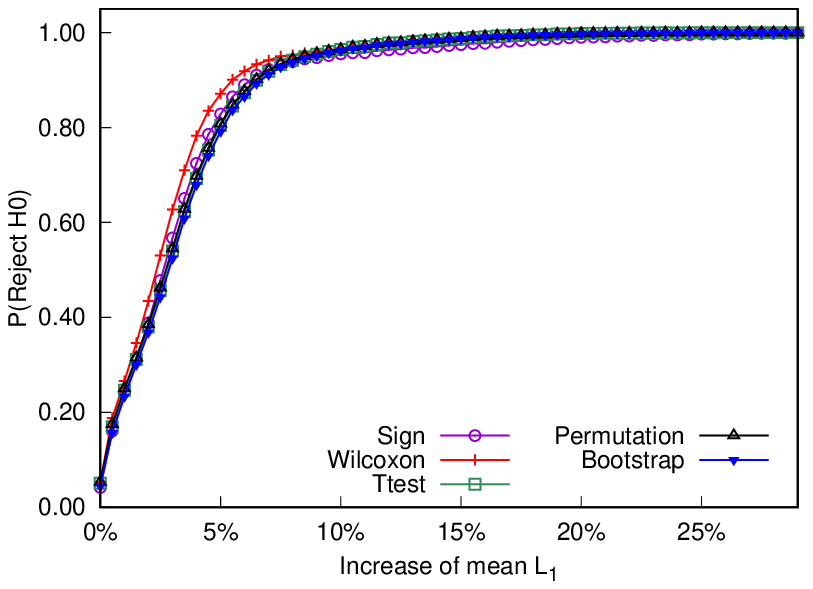}
		\subcaption{40 queries}
		\label{trec3q40}
	\end{subfigure}\\
	\begin{subfigure}[b]{0.49\textwidth}
		\includegraphics[width=\textwidth]{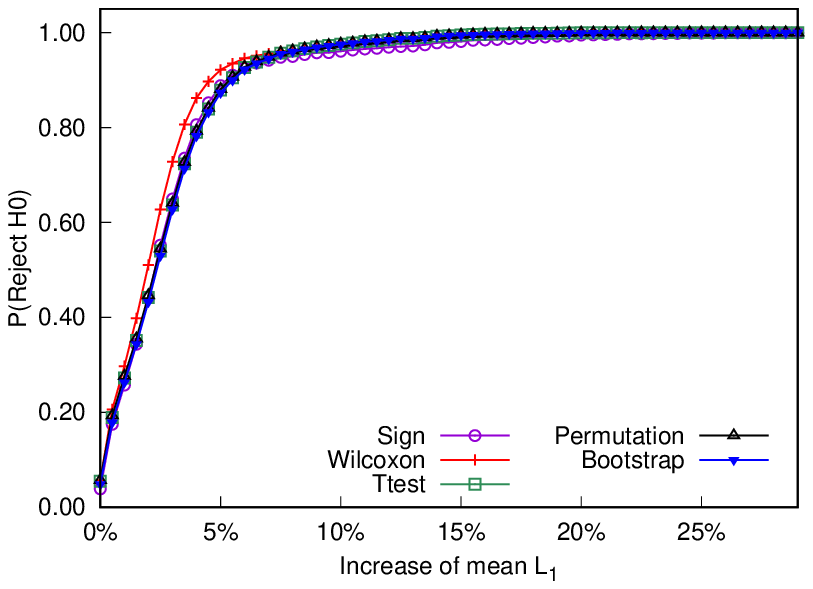}
		\subcaption{50 queries}
		\label{trec3q50}
	\end{subfigure}            
	\caption{Average $P(Reject\ H_0)$  ($\alpha=0.05$) in TREC 3. }
	\label{3figs}
\end{figure*}
\begin{figure*}
	\centering
	\begin{subfigure}[b]{0.49\textwidth}
		\includegraphics[width=\textwidth]{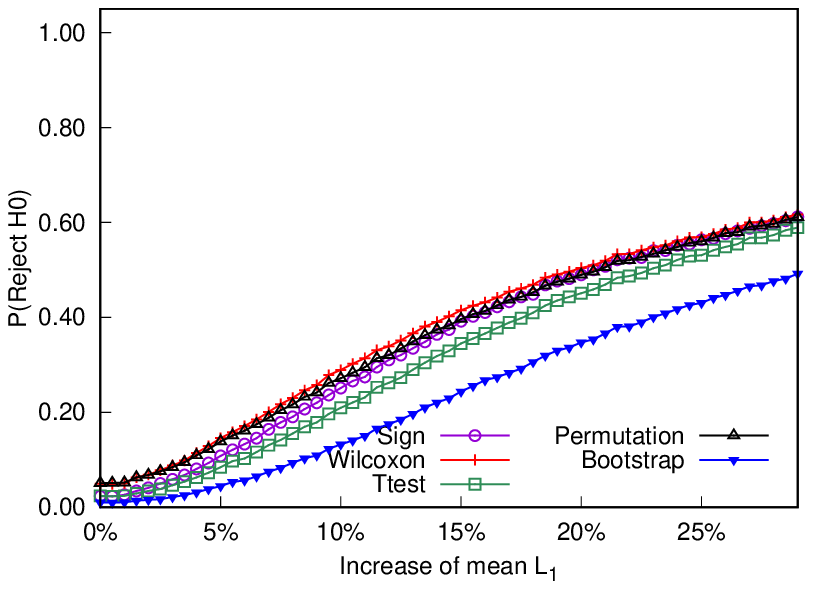}
		\subcaption{10 queries}
		\label{trec5q10}
	\end{subfigure}
	\begin{subfigure}[b]{0.49\textwidth}
		\includegraphics[width=\textwidth]{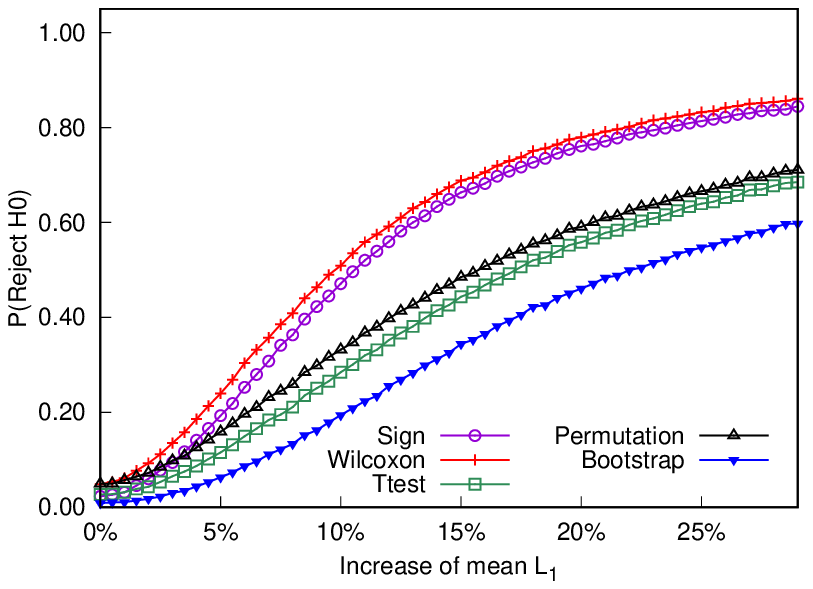}
		\subcaption{20 queries}
		\label{trec5q20}
	\end{subfigure}\\
	\begin{subfigure}[b]{0.49\textwidth}
		\includegraphics[width=\textwidth]{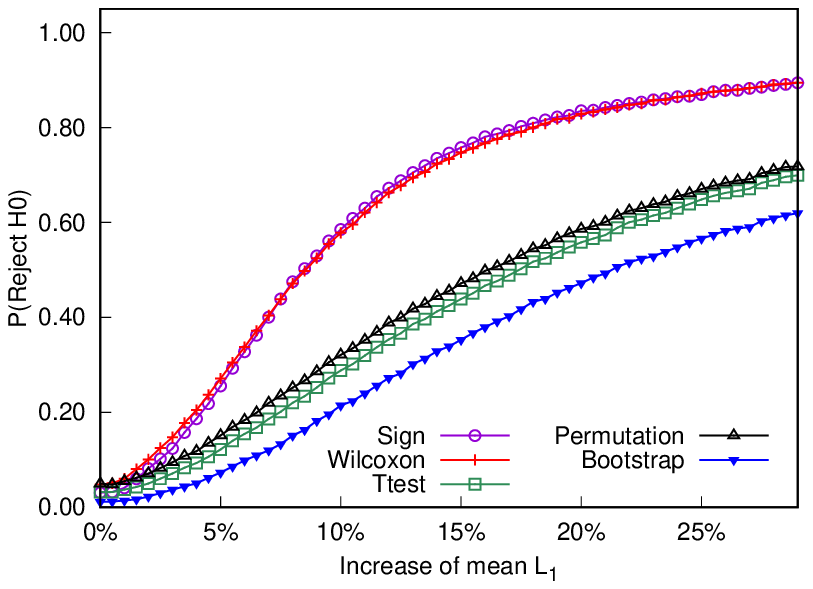}
		\subcaption{30 queries}
		\label{trec5q30}
	\end{subfigure}
	\begin{subfigure}[b]{0.49\textwidth}
		\includegraphics[width=\textwidth]{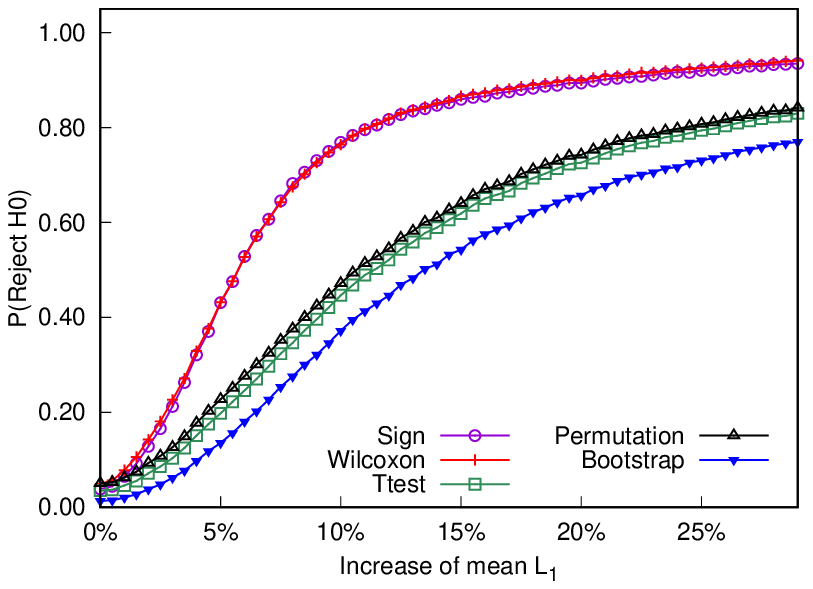}
		\subcaption{40 queries}
		\label{trec5q40}
	\end{subfigure}\\
	\begin{subfigure}[b]{0.49\textwidth}
		\includegraphics[width=\textwidth]{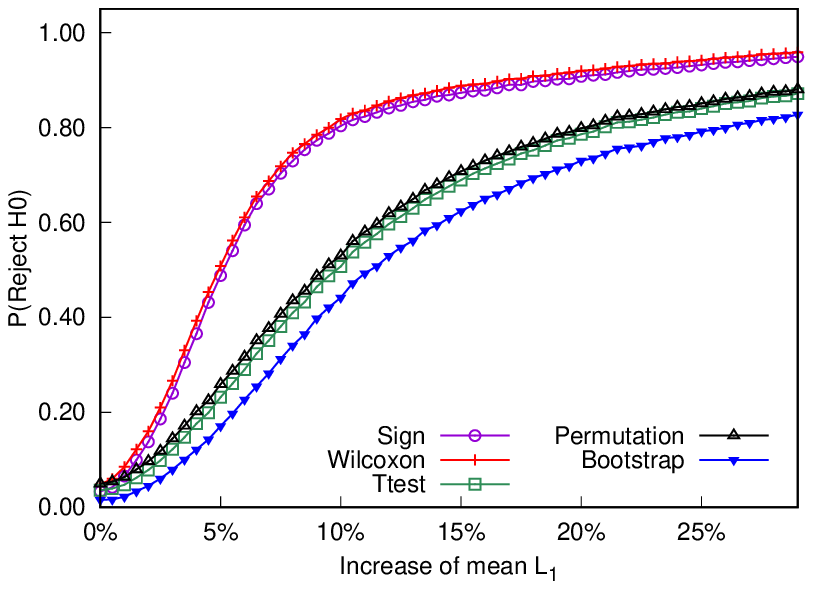}
		\subcaption{50 queries}
		\label{trec5q50}
	\end{subfigure}            
	\caption{Average $P(Reject\ H_0)$  ($\alpha=0.05$) in TREC 5.}
	\label{5figs}
\end{figure*}

\begin{figure*}
	\centering
	\begin{subfigure}[b]{0.49\textwidth}
		\includegraphics[width=\textwidth]{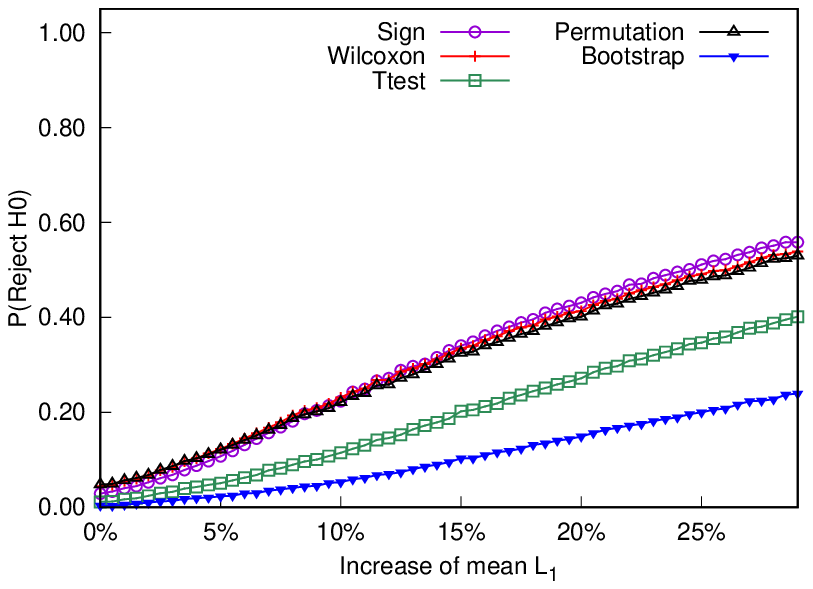}
		\subcaption{10 queries}
		\label{trec6q10}
	\end{subfigure}
	\begin{subfigure}[b]{0.49\textwidth}
		\includegraphics[width=\textwidth]{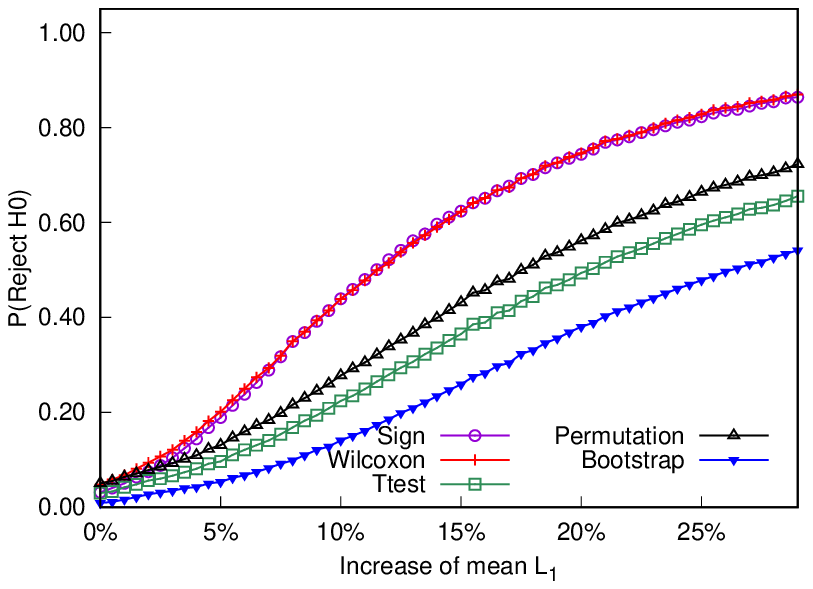}
		\subcaption{20 queries}
		\label{trec6q20}
	\end{subfigure}\\
	\begin{subfigure}[b]{0.49\textwidth}
		\includegraphics[width=\textwidth]{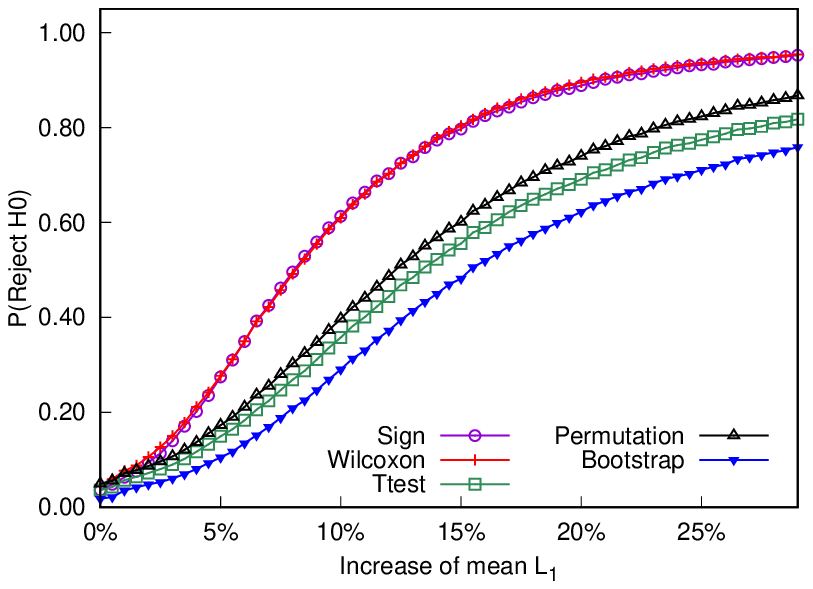}
		\subcaption{30 queries}
		\label{trec6q30}
	\end{subfigure}
	\begin{subfigure}[b]{0.49\textwidth}
		\includegraphics[width=\textwidth]{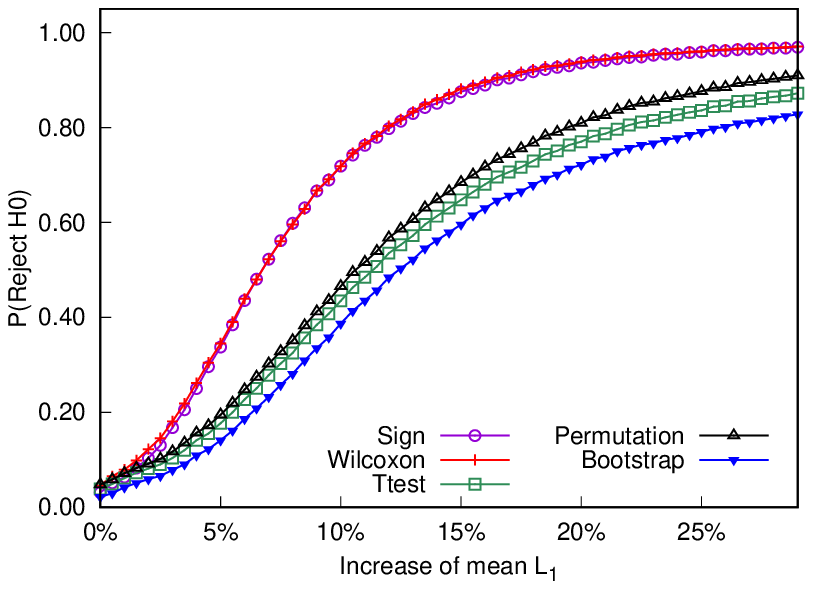}
		\subcaption{40 queries}
		\label{trec6q40}
	\end{subfigure}\\
	\begin{subfigure}[b]{0.49\textwidth}
		\includegraphics[width=\textwidth]{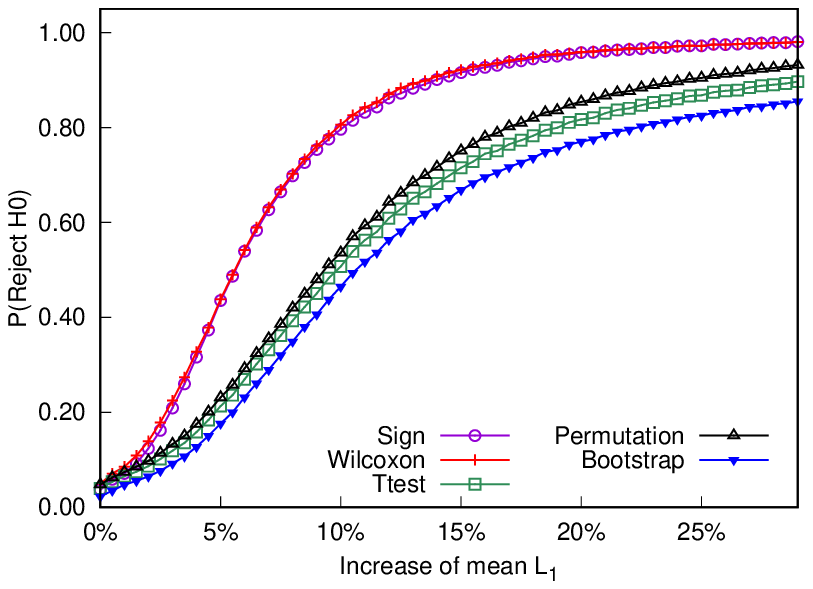}
		\subcaption{50 queries}
		\label{trec6q50}
	\end{subfigure}            
			\caption{Average $P(Reject\ H_0)$  ($\alpha=0.05$) in TREC 6.}
	\label{6figs}
\end{figure*}

\begin{figure*}
	\centering
	\begin{subfigure}[b]{0.49\textwidth}
		\includegraphics[width=\textwidth]{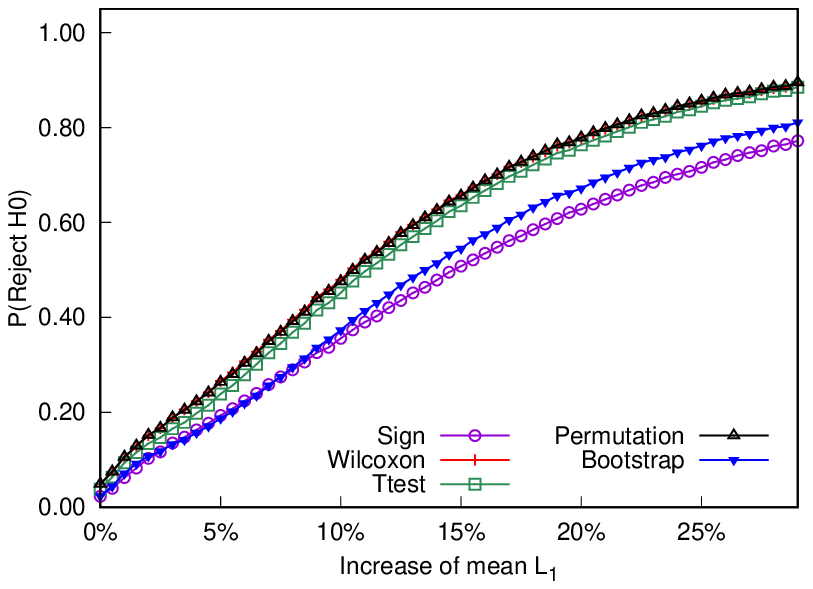}
		\subcaption{10 queries}
		\label{trec7q10}
	\end{subfigure}
	\begin{subfigure}[b]{0.49\textwidth}
		\includegraphics[width=\textwidth]{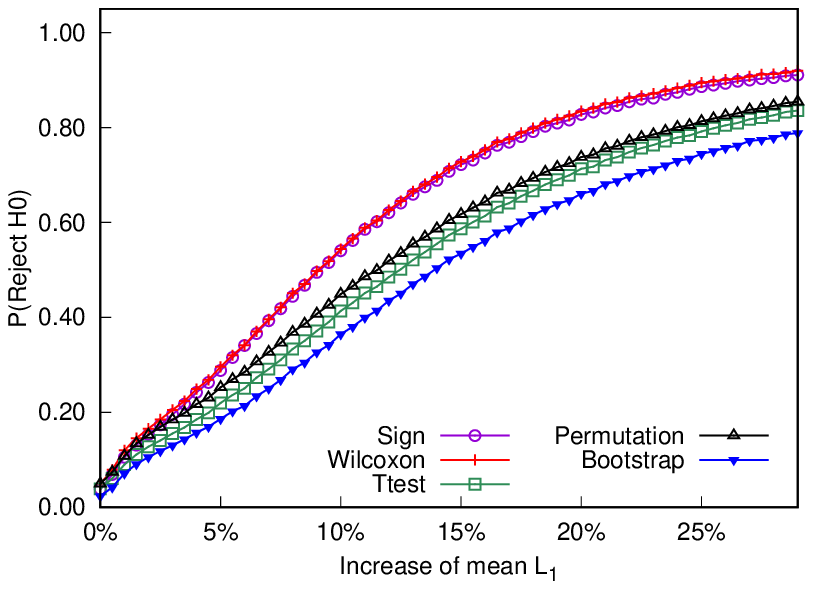}
		\subcaption{20 queries}
		\label{trec7q20}
	\end{subfigure}\\
	\begin{subfigure}[b]{0.49\textwidth}
		\includegraphics[width=\textwidth]{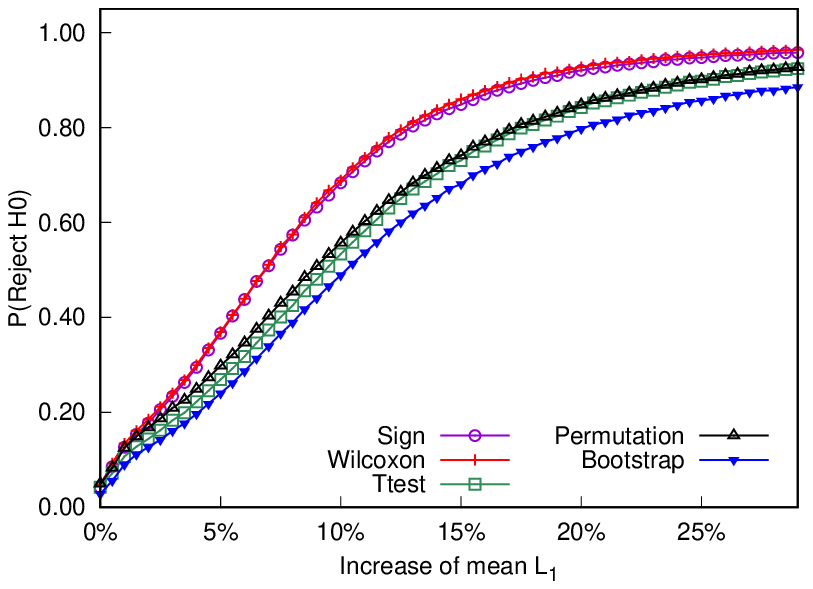}
		\subcaption{30 queries}
		\label{trec7q30}
	\end{subfigure}
	\begin{subfigure}[b]{0.49\textwidth}
		\includegraphics[width=\textwidth]{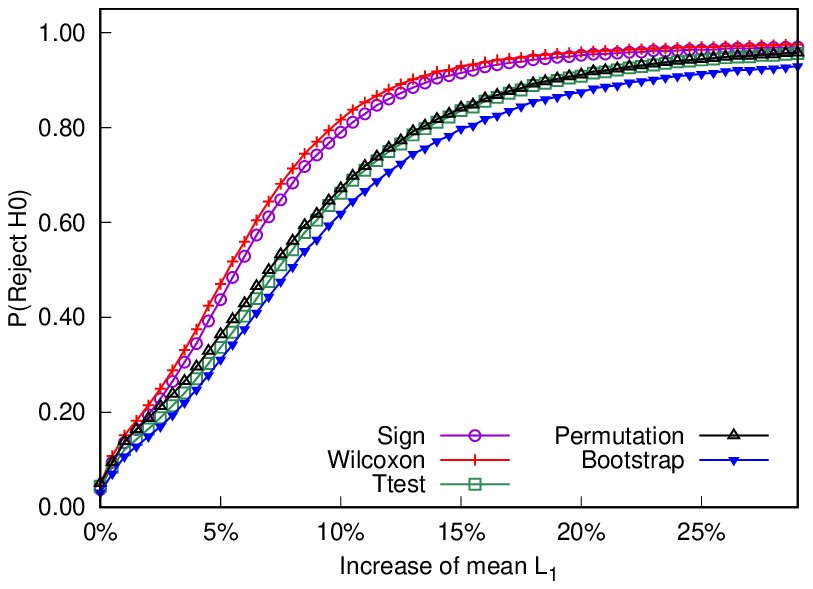}
		\subcaption{40 queries}
		\label{trec7q40}
	\end{subfigure}\\
	\begin{subfigure}[b]{0.49\textwidth}
		\includegraphics[width=\textwidth]{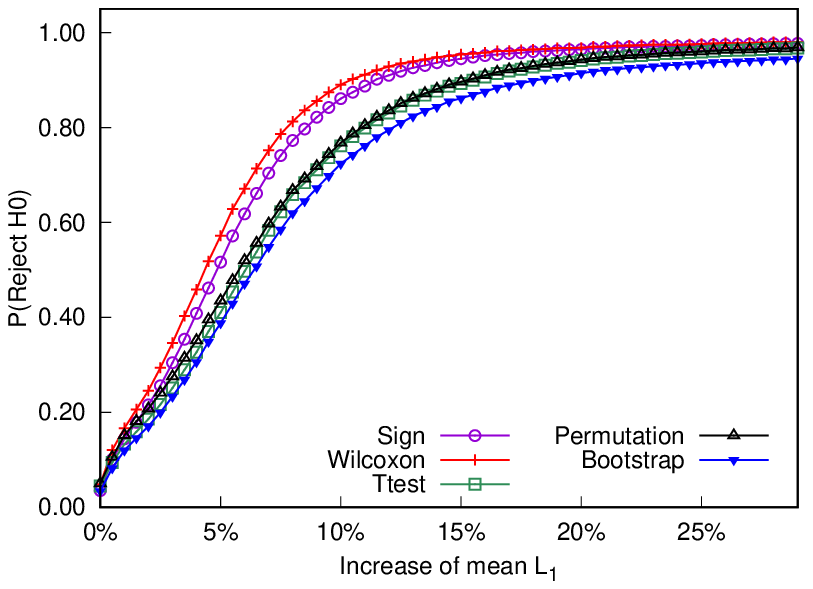}
		\subcaption{50 queries}
		\label{trec7q50}
	\end{subfigure}            
		\caption{Average $P(Reject\ H_0)$  ($\alpha=0.05$) in TREC 7.}	
	\label{7figs}
\end{figure*}

\begin{figure*}
	\centering
	\begin{subfigure}[b]{0.49\textwidth}
		\includegraphics[width=\textwidth]{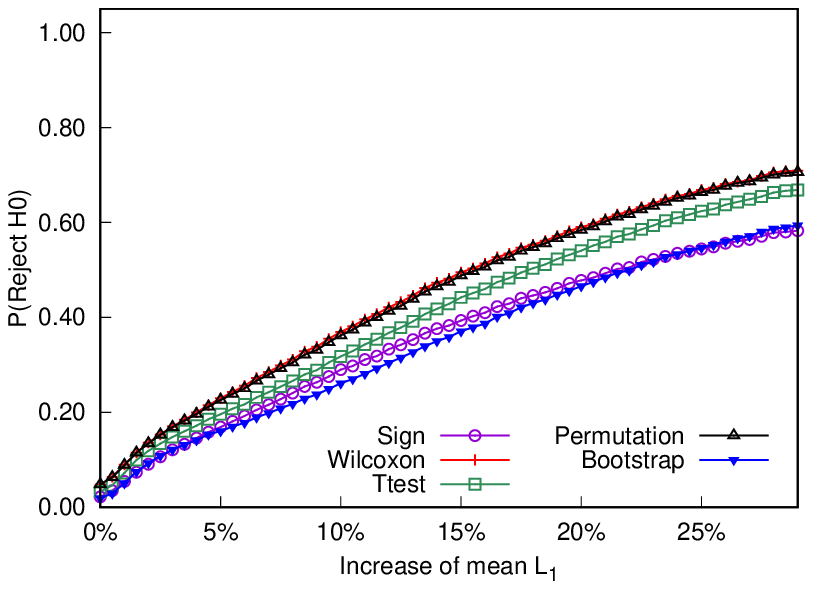}
		\subcaption{10 queries}
		\label{trec8q10}
	\end{subfigure}
	\begin{subfigure}[b]{0.49\textwidth}
		\includegraphics[width=\textwidth]{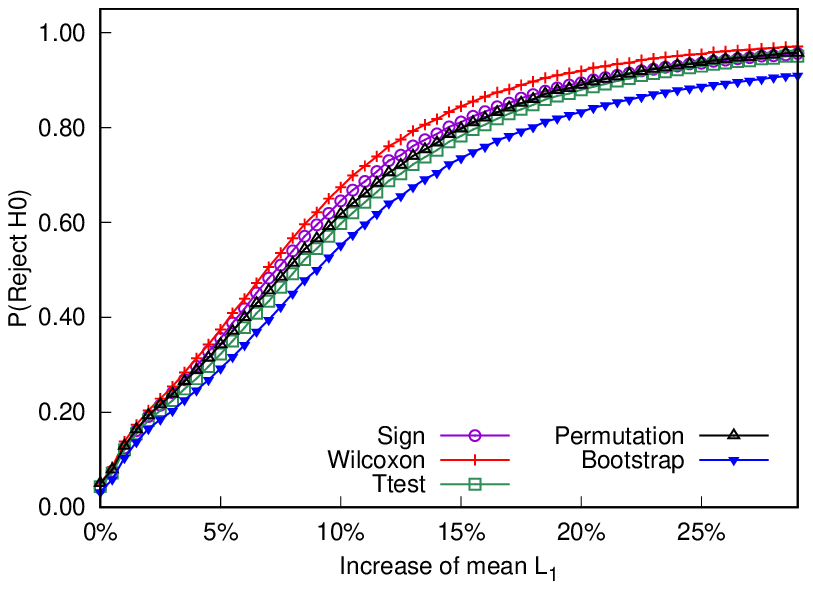}
		\subcaption{20 queries}
		\label{trec8q20}
	\end{subfigure}\\
	\begin{subfigure}[b]{0.49\textwidth}
		\includegraphics[width=\textwidth]{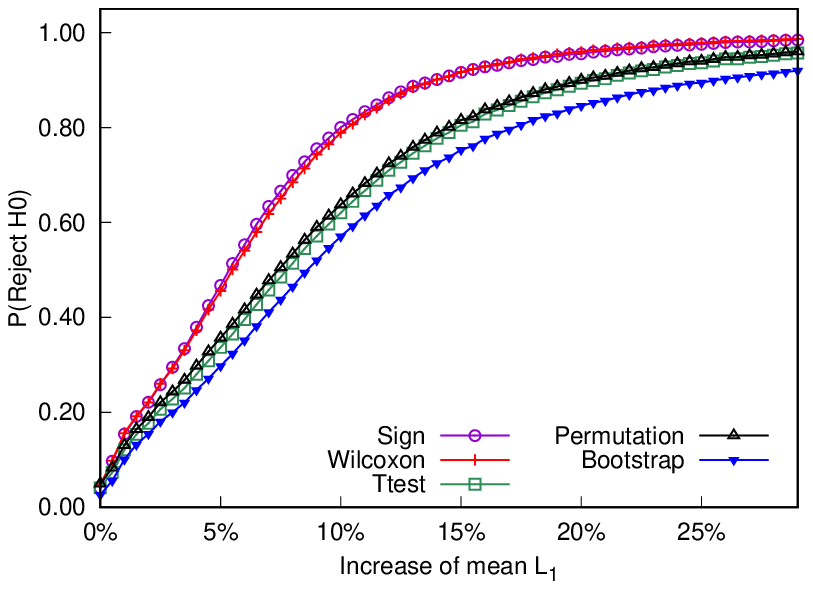}
		\subcaption{30 queries}
		\label{trec8q30}
	\end{subfigure}
	\begin{subfigure}[b]{0.49\textwidth}
		\includegraphics[width=\textwidth]{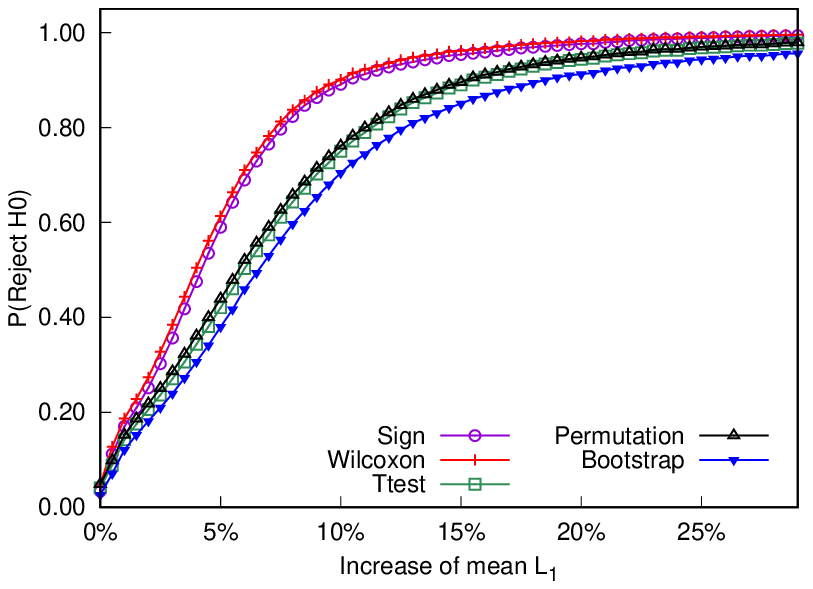}
		\subcaption{40 queries}
		\label{trec8q40}
	\end{subfigure}\\
	\begin{subfigure}[b]{0.49\textwidth}
		\includegraphics[width=\textwidth]{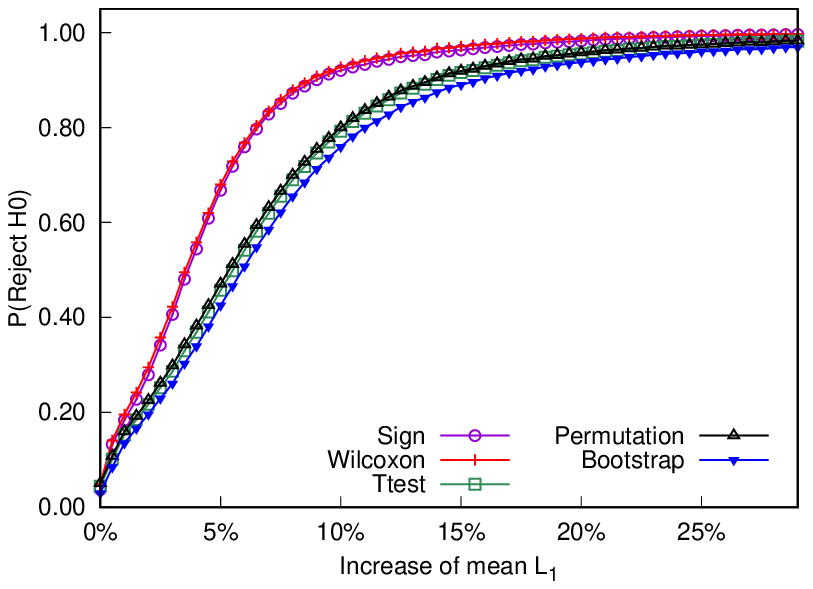}
		\subcaption{50 queries}
		\label{trec8q50}
	\end{subfigure}            
		\caption{Average $P(Reject\ H_0)$  ($\alpha=0.05$) in TREC 8.}
	\label{8figs}
\end{figure*}

Let us now analyse the tests when $H0$ is false. Each experiment compared one SD model against a transformed model, and we obtained different transformed models by increasing $\mu_1^{(i)}$ up to 30\% (in steps of 0.5\%). The resulting power curves, which plot the probability of rejection of the null hypothesis (averaged over all systems and samples), are shown in Figures \ref{3figs}, \ref{5figs}, \ref{6figs}, \ref{7figs} and \ref{8figs}. 

With less than 30 queries, all tests perform poorly (many type II errors). This result is in line with previous studies \cite{Webber:2008:SPR:1458082.1458158} and 
provides further empirical evidence on the need for large sets of queries 
when we want to derive statistical conclusions about the superiority of one system over the other. Many query splitting studies analysed significance
tests using less than 30 queries in each split and, therefore, their results must be taken with caution. 
Our discussion below therefore focuses on the graphs with more than 30 queries (Figures \ref{3figs}, \ref{5figs}, \ref{6figs}, \ref{7figs} and \ref{8figs}).

The form of each power curve is related to the difficulty of the track. 
For example, on average, TREC-3 has many relevant documents and, thus, the differences in performance between systems are higher and easier to detect by the tests.
As a consequence, TREC-3 plots tend to show a right angle. 

The power curves are quite revealing on the relative merits of the significance tests. 
The sign test and Wilcoxon perform the best at rejecting $H0$ when it is false. 
The permutation test, the t-test and bootstrap are inferior to both the sign test and Wilcoxon. 
Our results suggest that the permutation test and the t-test have a similar behavior (and bootstrap is slightly inferior to them). 
This outcome is in agreement with the study presented by Smucker et al. \cite{Smucker:2007}. 
But Smucker et al. did not analyse the power of the tests and, furthermore, they evaluated Wilcoxon and the sign test in terms of how well their results match with the results
yielded by the permutation test. As argued above, we believe that significance tests should be compared with no a priori assumptions about which test is the best.   

Our results also agree with accepted statistical principles. It is known that the power of a statistical test is mainly affected by: i) the effect size (the difference between the null and alternative values), ii) the sample size, iii) the variability in the samples, and iv) the significance level of the test. 
The power of a test depends on these four factors, and it is not uncommon that simpler tests perform better than complex ones. Furthermore, the 
relative merits of the tests change under different conditions. For example, our TREC8 experiments show that when the sample size is small (10 queries) and the effect size is high (more than 15\%) the sign test has less power. Our results also agree with the findings reported by Conover about the relative loss of power of the permutation test (see Section 5.11, \cite{conover99}). 
As a matter of fact, simulation studies conducted by Conover and Iman suggested a preference on the tests (t-test $<<$ permutation $<<$ Wilcoxon) that matches with ours. 
Our study also agrees with Kempthorne and Doerfler \cite{Kempthorne}, who concluded that the permutation test behaves very well under $H0$. 
The permutation test matches very well with the significance value (see Fig. \ref{typeIfigs}). However, only Wilcoxon behaves well under both $H0$ (Fig. \ref{typeIfigs}) and $H1$ (power curves).

\subsection{Discussion}
\label{s:validity}

\begin{figure}[t]
	\centering
	
	\includegraphics[width=0.8\columnwidth]{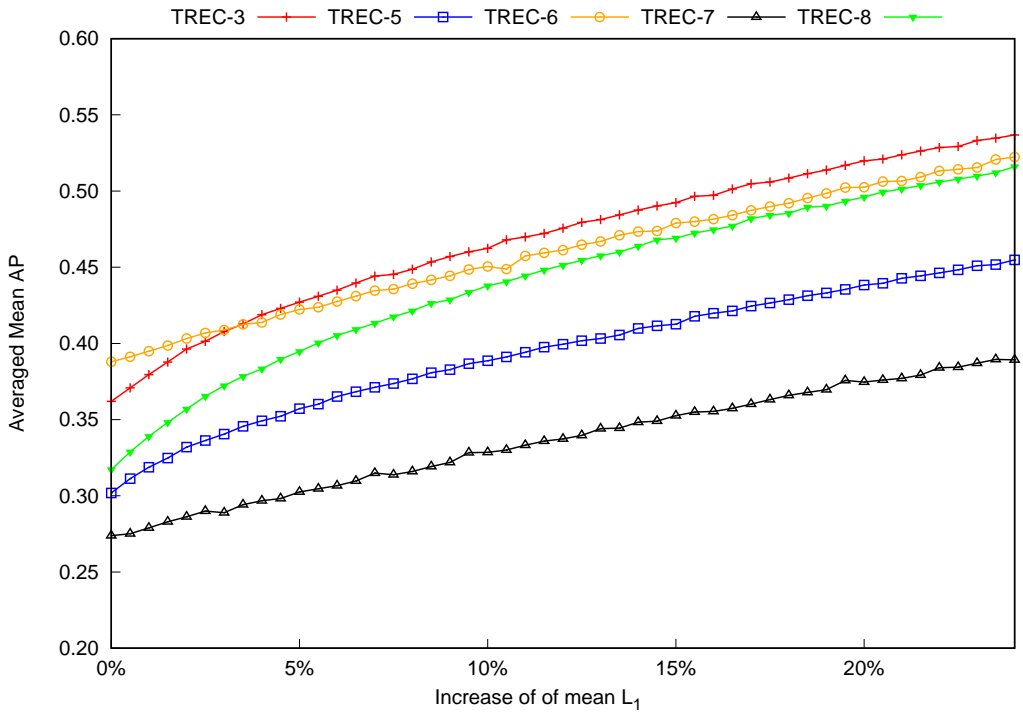}
	\caption{Average Mean Average Precision against increased means of the $L_1$ distributions.}
	\label{f:increasing}
\end{figure}

Studying significance tests based on the APs obtained from the simulated runs is a reliable way to understand
how the tests detect differences in search performance. An essential component of our methodology is the way in which we produce SD models with increased performance. 
In section \ref{astsdm}, we claimed that increasing $\mu_1^{(i)}$ leads to better performance. 
Figure \ref{f:increasing} plots the AP --averaged over all systems and running 50 simulations per model-- with varying increases of $\mu_1^{(i)}$.
The curves, which are monotonic increasing, demonstrate that manipulating the mean of $L_1$ leads to models with increasingly better performance. 

\begin{figure*}
	\centering
		\includegraphics[width=\textwidth]{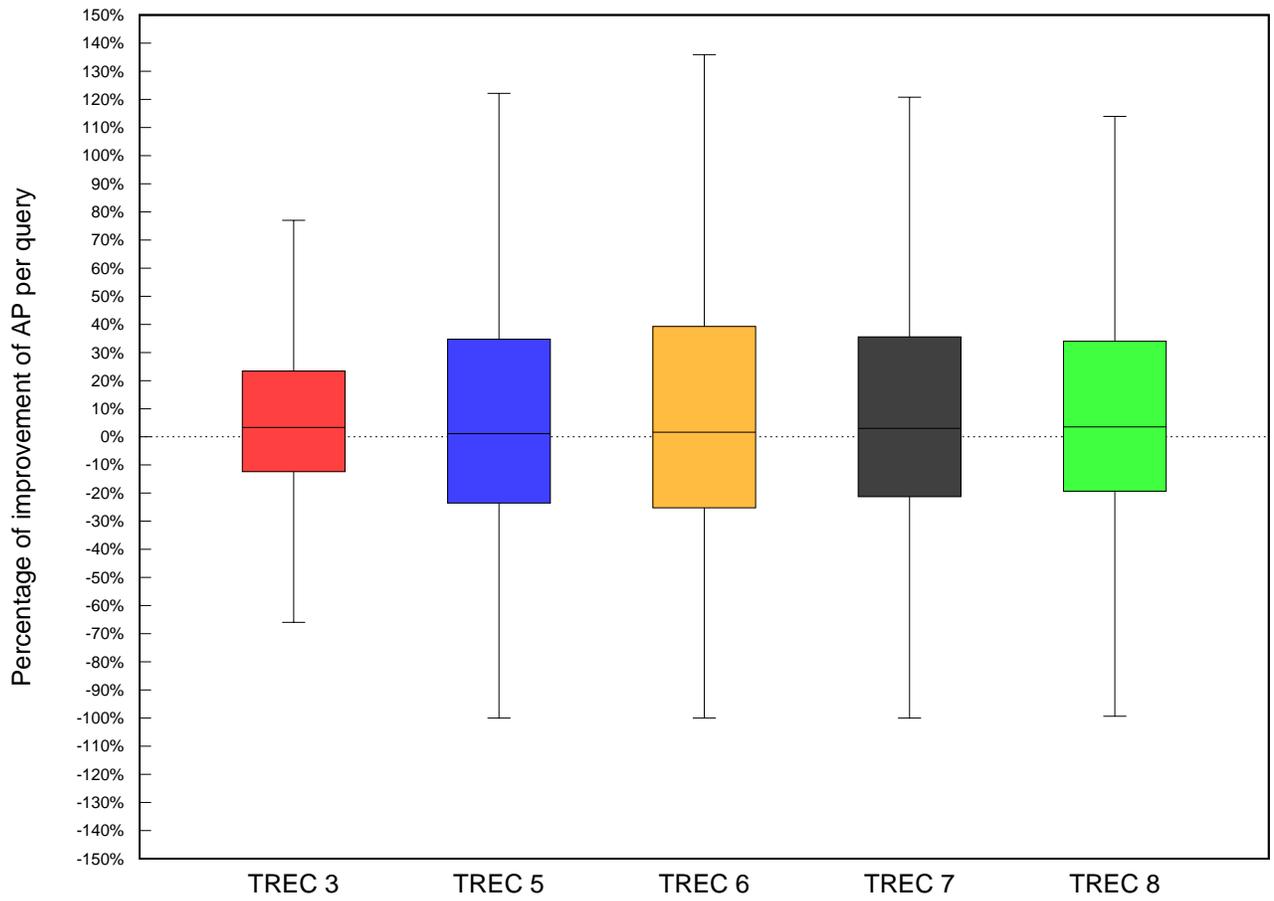}
	\caption{Effect of a 5\% increase in the mean of $L_1$. For each collection, the boxplot summarises the distribution of improvements across queries.}
	\label{boxplotnew}
\end{figure*}

Observe that increasing $\mu_1^{(i)}$ does not improve the performance of the modeled system for every query. 
The process is stochastic in nature and the improved systems tend to perform better 
for many queries, but they also lead to decreased performance 
for a few queries. This is a natural consequence of the sampling process (sampling
from a better model tends to produce better performance, but individual samples show some variance). 
To further illustrate this point, Figure \ref{boxplotnew} shows the effect of a 5\% increment in 
$\mu_1^{(i)}$ (similar boxplots were obtained for the other percentages of improvements). The figure
presents a summary of the distribution of $\Delta AP$ (\% variation: AP improved model vs AP original model) for all system-query pairs in our simulated study\footnote{The boxplot represents 100 simulations for each query-system pair.}. In all collections, the median improvement in AP is positive, but there is a large variance, with many queries improved and some other queries damaged. 
This demonstrates that the simulation is realistic (the system that is intrinsically better underperforms for some individual queries) and reflects the typical situation of topic variation in IR experiments
(an improvement in search technology leads to improved performance for the majority of topics, but
it also leads to poorer performance for a minority of topics).

\section{Conclusions}

In this work, we have thoroughly analysed a number of significance tests that have been commonly employed in IR experiments. 
A crucial contribution of our paper is the proposal of an innovative methodology to compare significance tests. Our method models simulated search systems and evaluates
the tests under complete certainty about the truth value of the null hypothesis. Following the lessons learnt in the area of Score
Distributions for IR, we built models --learnt from TREC runs--- that mimic the behavior of real search systems, we created different
situations where the null hypothesis is true or false, and we evaluated the ability of significance tests to correctly accept or reject the null hypothesis.

The experiments performed revealed that Wilcoxon and the sign test are the most reliable tests for IR evaluation. Both of them have more power
than the permutation test, bootstrap and the t-test. Furthermore, the behavior of Wilcoxon and the sign test concerning type I errors is solid. 
Previous studies \cite{Smucker:2007} claimed that the use of Wilcoxon and sign test should be discontinued. Our study reveals otherwise. We confirmed
that the t-test and bootstrap resemble more the permutation test, but we also provided substantive evidence on the weak power of the permutation test when compared
with Wilcoxon or the sign test. Smucker et al \cite{Smucker:2007} showed that, in comparison with the permutation test, Wilcoxon and the sign test produce different p-values. Such outcome
led the authors to argue against the two non-parametric tests (in a study where miss and false alarm rates were computed based on the agreement between each test and the permutation test). 
Our methodology is agnostic about which test is more reliable, adds another valuable tool to IR practitioners, and shows that the current support to the t-test or permutation should be revisited. 
Following our study, we would recommend IR practitioners to employ Wilcoxon or the sign test. Compared to the three other tests, 
Wilcoxon and the sign tests have more power and, thus, a higher ability to detect real improvements in search technologies. 
This ability is instrumental in advancing the field of IR. 

Our results are also in agreement with well-known statistical facts. We showed that the larger the query set is, the more reliable the comparison is. 
Experiments with less than 30 queries are questionable and, therefore, it is misleading to compare significance tests based on query-splitting methods whose splits have less
than 30 queries. 
On the other hand, Conover \cite{conover99} in his authoritative book on non-parametric statistics, exposed a number of cases where 
parametric methods, such as the t-test, have low power compared to non-parametric methods based on ranks. Such behavior is particularly apparent
with data that show specific deviations from the normal distribution. 
Our evaluation fits well with such statistical knowledge (as a matter of fact, the distribution of differences in ad-hoc retrieval is known not to follow a normal distribution).
In his book, Conover also referred to simulation studies where the permutation test showed a relative lack of power when compared to tests based on ranks. 
Overall, our empirical findings are in line with Conover's findings. In a thorough series of experiments performed with multiple retrieval collections,
we showed that the t-test and the permutation test --whose p-values are similar to those of the t-test-- have less power than the sign test and Wilcoxon. This suggests that, 
under the typical conditions of IR evaluation, we should prefer rank-based methods over alternative significance tests. 
Finally, the weak results obtained with the bootstrap test were a bit surprising. Such poor behavior suggests the need for more effective bootstrap methods in IR experimentation.

\section*{Acknowledgements}

This work has received financial support from the i) ``Ministerio de Economía
y Competitividad'' of the Government of Spain and FEDER
Funds under the research project TIN2015-64282-R, ii) Xunta de Galicia (project GPC 2016/035),
and iii) Xunta de Galicia – ``Consellería de Cultura, Educación e Ordenación Universitaria'' 
and the European Regional Development Fund (ERDF) through the following 2016-2019 accreditations:
ED431G/01 (``Centro singular de investigación de Galicia'') and ED431G/08.

We also thank the anonymous reviewers for their really useful suggestions and comments.  

\bibliographystyle{apacite}
\bibliography{data_base}

\begin{thebibliography}{}

\bibitem [\protect \citeauthoryear {%
Arampatzis%
, Beney%
, Koster%
\BCBL {}\ \BBA {} van~der Weide%
}{%
Arampatzis%
\ \protect \BOthers {.}}{%
{\protect \APACyear {2000}}%
}]{%
Arampatzis2000}
\APACinsertmetastar {%
Arampatzis2000}%
\begin{APACrefauthors}%
Arampatzis, A.%
, Beney, J.%
, Koster, C\BPBI H\BPBI A.%
\BCBL {}\ \BBA {} van~der Weide, T\BPBI P.%
\end{APACrefauthors}%
\unskip\
\newblock
\APACrefYearMonthDay{2000}{}{}.
\newblock
{\BBOQ}\APACrefatitle {{Incrementality, Half-life, and Threshold Optimization
  for Adaptive Document Filtering.}} {{Incrementality, Half-life, and Threshold
  Optimization for Adaptive Document Filtering.}}{\BBCQ}
\newblock
\BIn{} \APACrefbtitle {Proceedings of the 9th Text REtrieval Conference, {TREC}
  2000.} {Proceedings of the 9th text retrieval conference, {TREC} 2000.}
\newblock
\APACaddressPublisher{}{National Institute for Science and Technology (NIST)}.
\PrintBackRefs{\CurrentBib}

\bibitem [\protect \citeauthoryear {%
Arampatzis%
, Kamps%
\BCBL {}\ \BBA {} Robertson%
}{%
Arampatzis%
, Kamps%
\BCBL {}\ \BBA {} Robertson%
}{%
{\protect \APACyear {2009}}%
}]{%
a2009}
\APACinsertmetastar {%
a2009}%
\begin{APACrefauthors}%
Arampatzis, A.%
, Kamps, J.%
\BCBL {}\ \BBA {} Robertson, S.%
\end{APACrefauthors}%
\unskip\
\newblock
\APACrefYearMonthDay{2009}{}{}.
\newblock
{\BBOQ}\APACrefatitle {{Where to stop reading a ranked list?}} {{Where to stop
  reading a ranked list?}}{\BBCQ}
\newblock
\BIn{} \APACrefbtitle {Proc. {SIGIR} '09} {Proc. {SIGIR} '09}\ (\BPGS\
  524--531).
\newblock
\APACaddressPublisher{New York, USA}{ACM Press}.
\PrintBackRefs{\CurrentBib}

\bibitem [\protect \citeauthoryear {%
Arampatzis%
\ \BBA {} Robertson%
}{%
Arampatzis%
\ \BBA {} Robertson%
}{%
{\protect \APACyear {2011}}%
}]{%
Arampatzis:2011}
\APACinsertmetastar {%
Arampatzis:2011}%
\begin{APACrefauthors}%
Arampatzis, A.%
\BCBT {}\ \BBA {} Robertson, S.%
\end{APACrefauthors}%
\unskip\
\newblock
\APACrefYearMonthDay{2011}{{\APACmonth{02}}}{}.
\newblock
{\BBOQ}\APACrefatitle {Modeling Score Distributions in Information Retrieval}
  {Modeling score distributions in information retrieval}.{\BBCQ}
\newblock
\APACjournalVolNumPages{Inf. Retr.}{14}{1}{26--46}.
\PrintBackRefs{\CurrentBib}

\bibitem [\protect \citeauthoryear {%
Arampatzis%
, Robertson%
\BCBL {}\ \BBA {} Kamps%
}{%
Arampatzis%
, Robertson%
\BCBL {}\ \BBA {} Kamps%
}{%
{\protect \APACyear {2009}}%
}]{%
Arampatzis2009}
\APACinsertmetastar {%
Arampatzis2009}%
\begin{APACrefauthors}%
Arampatzis, A.%
, Robertson, S.%
\BCBL {}\ \BBA {} Kamps, J.%
\end{APACrefauthors}%
\unskip\
\newblock
\APACrefYearMonthDay{2009}{}{}.
\newblock
{\BBOQ}\APACrefatitle {{Score Distributions in Information Retrieval}} {{Score
  Distributions in Information Retrieval}}.{\BBCQ}
\newblock
\BIn{} \APACrefbtitle {Proceedings {ICTIR} 2009} {Proceedings {ICTIR} 2009}\
  (\BPGS\ 139--151).
\PrintBackRefs{\CurrentBib}

\bibitem [\protect \citeauthoryear {%
Arampatzis%
, Zagoris%
\BCBL {}\ \BBA {} Chatzichristofis%
}{%
Arampatzis%
\ \protect \BOthers {.}}{%
{\protect \APACyear {2013}}%
}]{%
ipm1}
\APACinsertmetastar {%
ipm1}%
\begin{APACrefauthors}%
Arampatzis, A.%
, Zagoris, K.%
\BCBL {}\ \BBA {} Chatzichristofis, S\BPBI A.%
\end{APACrefauthors}%
\unskip\
\newblock
\APACrefYearMonthDay{2013}{{\APACmonth{01}}}{}.
\newblock
{\BBOQ}\APACrefatitle {{Dynamic two-stage image retrieval from large multimedia
  databases}} {{Dynamic two-stage image retrieval from large multimedia
  databases}}.{\BBCQ}
\newblock
\APACjournalVolNumPages{Information Processing \& Management}{49}{1}{274-285}.
\PrintBackRefs{\CurrentBib}

\bibitem [\protect \citeauthoryear {%
Baumgarten%
}{%
Baumgarten%
}{%
{\protect \APACyear {1999}}%
}]{%
Baumgarten:1999}
\APACinsertmetastar {%
Baumgarten:1999}%
\begin{APACrefauthors}%
Baumgarten, C.%
\end{APACrefauthors}%
\unskip\
\newblock
\APACrefYearMonthDay{1999}{{\APACmonth{08}}}{}.
\newblock
{\BBOQ}\APACrefatitle {{A probabilistic solution to the selection and fusion
  problem in distributed information retrieval}} {{A probabilistic solution to
  the selection and fusion problem in distributed information
  retrieval}}.{\BBCQ}
\newblock
\BIn{} \APACrefbtitle {Proceedings {SIGIR} '99} {Proceedings {SIGIR} '99}\
  (\BPGS\ 246--253).
\newblock
\APACaddressPublisher{New York, USA}{ACM Press}.
\PrintBackRefs{\CurrentBib}

\bibitem [\protect \citeauthoryear {%
Bookstein%
}{%
Bookstein%
}{%
{\protect \APACyear {1977}}%
}]{%
Bookstein1977}
\APACinsertmetastar {%
Bookstein1977}%
\begin{APACrefauthors}%
Bookstein, A.%
\end{APACrefauthors}%
\unskip\
\newblock
\APACrefYearMonthDay{1977}{{\APACmonth{01}}}{}.
\newblock
{\BBOQ}\APACrefatitle {{When the most "pertinent" document should not be
  retrieved An analysis of the Swets model}} {{When the most "pertinent"
  document should not be retrieved An analysis of the Swets model}}.{\BBCQ}
\newblock
\APACjournalVolNumPages{Information Processing \& Management}{13}{6}{377--383}.
\PrintBackRefs{\CurrentBib}

\bibitem [\protect \citeauthoryear {%
Conover%
}{%
Conover%
}{%
{\protect \APACyear {1999}}%
}]{%
conover99}
\APACinsertmetastar {%
conover99}%
\begin{APACrefauthors}%
Conover, W.%
\end{APACrefauthors}%
\unskip\
\newblock
\APACrefYear{1999}.
\newblock
\APACrefbtitle {Practical nonparametric statistics} {Practical nonparametric
  statistics}\ (\PrintOrdinal{3rd}\ \BEd).
\newblock
\APACaddressPublisher{New York}{Wiley}.
\PrintBackRefs{\CurrentBib}

\bibitem [\protect \citeauthoryear {%
Cormack%
\ \BBA {} Lynam%
}{%
Cormack%
\ \BBA {} Lynam%
}{%
{\protect \APACyear {2006}}%
}]{%
Cormack:2006}
\APACinsertmetastar {%
Cormack:2006}%
\begin{APACrefauthors}%
Cormack, G\BPBI V.%
\BCBT {}\ \BBA {} Lynam, T\BPBI R.%
\end{APACrefauthors}%
\unskip\
\newblock
\APACrefYearMonthDay{2006}{}{}.
\newblock
{\BBOQ}\APACrefatitle {Statistical Precision of Information Retrieval
  Evaluation} {Statistical precision of information retrieval
  evaluation}.{\BBCQ}
\newblock
\BIn{} \APACrefbtitle {Proceedings {SIGIR} 2006} {Proceedings {SIGIR} 2006}\
  (\BPGS\ 533--540).
\newblock
\APACaddressPublisher{New York, NY, USA}{ACM}.
\PrintBackRefs{\CurrentBib}

\bibitem [\protect \citeauthoryear {%
Cormack%
\ \BBA {} Lynam%
}{%
Cormack%
\ \BBA {} Lynam%
}{%
{\protect \APACyear {2007}}%
}]{%
Cormack2007}
\APACinsertmetastar {%
Cormack2007}%
\begin{APACrefauthors}%
Cormack, G\BPBI V.%
\BCBT {}\ \BBA {} Lynam, T\BPBI R.%
\end{APACrefauthors}%
\unskip\
\newblock
\APACrefYearMonthDay{2007}{{\APACmonth{07}}}{}.
\newblock
{\BBOQ}\APACrefatitle {{Validity and power of t-test for comparing MAP and
  GMAP}} {{Validity and power of t-test for comparing MAP and GMAP}}.{\BBCQ}
\newblock
\BIn{} \APACrefbtitle {Proceedings {SIGIR} '07} {Proceedings {SIGIR} '07}\
  (\BPG~753-754).
\newblock
\APACaddressPublisher{New York, USA}{ACM Press}.
\PrintBackRefs{\CurrentBib}

\bibitem [\protect \citeauthoryear {%
Cummins%
}{%
Cummins%
}{%
{\protect \APACyear {2011}}%
}]{%
Cummins2011}
\APACinsertmetastar {%
Cummins2011}%
\begin{APACrefauthors}%
Cummins, R.%
\end{APACrefauthors}%
\unskip\
\newblock
\APACrefYearMonthDay{2011}{}{}.
\newblock
{\BBOQ}\APACrefatitle {Measuring the Ability of Score Distributions to Model
  Relevance} {Measuring the ability of score distributions to model
  relevance}.{\BBCQ}
\newblock
\BIn{} \APACrefbtitle {Proceedings of the 7th Asia Conference on Information
  Retrieval Technology} {Proceedings of the 7th asia conference on information
  retrieval technology}\ (\BPGS\ 25--36).
\newblock
\APACaddressPublisher{Berlin, Heidelberg}{Springer-Verlag}.
\PrintBackRefs{\CurrentBib}

\bibitem [\protect \citeauthoryear {%
Cummins%
}{%
Cummins%
}{%
{\protect \APACyear {2014}}%
}]{%
Cummins:2014:DSD:2576772.2559170}
\APACinsertmetastar {%
Cummins:2014:DSD:2576772.2559170}%
\begin{APACrefauthors}%
Cummins, R.%
\end{APACrefauthors}%
\unskip\
\newblock
\APACrefYearMonthDay{2014}{}{}.
\newblock
{\BBOQ}\APACrefatitle {Document Score Distribution Models for Query Performance
  Inference and Prediction} {Document score distribution models for query
  performance inference and prediction}.{\BBCQ}
\newblock
\APACjournalVolNumPages{ACM Trans. Inf. Syst.}{32}{1}{1--28}.
\PrintBackRefs{\CurrentBib}

\bibitem [\protect \citeauthoryear {%
Cummins%
\ \BBA {} O'Riordan%
}{%
Cummins%
\ \BBA {} O'Riordan%
}{%
{\protect \APACyear {2012}}%
}]{%
Cummins:2012:TVS:2260641.2260691}
\APACinsertmetastar {%
Cummins:2012:TVS:2260641.2260691}%
\begin{APACrefauthors}%
Cummins, R.%
\BCBT {}\ \BBA {} O'Riordan, C.%
\end{APACrefauthors}%
\unskip\
\newblock
\APACrefYearMonthDay{2012}{}{}.
\newblock
{\BBOQ}\APACrefatitle {On Theoretically Valid Score Distributions in
  Information Retrieval} {On theoretically valid score distributions in
  information retrieval}.{\BBCQ}
\newblock
\BIn{} \APACrefbtitle {Proceedings of {ECIR} 2012} {Proceedings of {ECIR}
  2012}\ (\BPGS\ 451--454).
\newblock
\APACaddressPublisher{Berlin, Heidelberg}{Springer-Verlag}.
\PrintBackRefs{\CurrentBib}

\bibitem [\protect \citeauthoryear {%
Dai%
, Kanoulas%
, Pavlu%
\BCBL {}\ \BBA {} Aslam%
}{%
Dai%
\ \protect \BOthers {.}}{%
{\protect \APACyear {2011}}%
}]{%
dai2011variational}
\APACinsertmetastar {%
dai2011variational}%
\begin{APACrefauthors}%
Dai, K.%
, Kanoulas, E.%
, Pavlu, V.%
\BCBL {}\ \BBA {} Aslam, J\BPBI A.%
\end{APACrefauthors}%
\unskip\
\newblock
\APACrefYearMonthDay{2011}{}{}.
\newblock
{\BBOQ}\APACrefatitle {Variational {Bayes} for modeling score distributions}
  {Variational {Bayes} for modeling score distributions}.{\BBCQ}
\newblock
\APACjournalVolNumPages{Information retrieval}{14}{1}{47--67}.
\PrintBackRefs{\CurrentBib}

\bibitem [\protect \citeauthoryear {%
Dempster%
, Laird%
\BCBL {}\ \BBA {} Rubin%
}{%
Dempster%
\ \protect \BOthers {.}}{%
{\protect \APACyear {1977}}%
}]{%
Dempster77maximumlikelihood}
\APACinsertmetastar {%
Dempster77maximumlikelihood}%
\begin{APACrefauthors}%
Dempster, A\BPBI P.%
, Laird, N\BPBI M.%
\BCBL {}\ \BBA {} Rubin, D\BPBI B.%
\end{APACrefauthors}%
\unskip\
\newblock
\APACrefYearMonthDay{1977}{}{}.
\newblock
{\BBOQ}\APACrefatitle {Maximum likelihood from incomplete data via the {EM}
  algorithm} {Maximum likelihood from incomplete data via the {EM}
  algorithm}.{\BBCQ}
\newblock
\APACjournalVolNumPages{Journal of the Royal Statistical
  Society}{39}{1}{1--38}.
\PrintBackRefs{\CurrentBib}

\bibitem [\protect \citeauthoryear {%
Efron%
\ \BBA {} Tibshirani%
}{%
Efron%
\ \BBA {} Tibshirani%
}{%
{\protect \APACyear {1993}}%
}]{%
Efron1993}
\APACinsertmetastar {%
Efron1993}%
\begin{APACrefauthors}%
Efron, B.%
\BCBT {}\ \BBA {} Tibshirani, R.%
\end{APACrefauthors}%
\unskip\
\newblock
\APACrefYear{1993}.
\newblock
\APACrefbtitle {{An Introduction to the Bootstrap}} {{An Introduction to the
  Bootstrap}}\ [Miscellaneous].
\newblock
\APACaddressPublisher{}{Macmillan Publishers Limited.}
\PrintBackRefs{\CurrentBib}

\bibitem [\protect \citeauthoryear {%
Hull%
}{%
Hull%
}{%
{\protect \APACyear {1993}}%
}]{%
Hull:1993}
\APACinsertmetastar {%
Hull:1993}%
\begin{APACrefauthors}%
Hull, D.%
\end{APACrefauthors}%
\unskip\
\newblock
\APACrefYearMonthDay{1993}{}{}.
\newblock
{\BBOQ}\APACrefatitle {Using Statistical Testing in the Evaluation of Retrieval
  Experiments} {Using statistical testing in the evaluation of retrieval
  experiments}.{\BBCQ}
\newblock
\BIn{} \APACrefbtitle {Proceedings {SIGIR} 1993} {Proceedings {SIGIR} 1993}\
  (\BPGS\ 329--338).
\newblock
\APACaddressPublisher{New York, NY, USA}{ACM}.
\PrintBackRefs{\CurrentBib}

\bibitem [\protect \citeauthoryear {%
Kanoulas%
, Dai%
, Pavlu%
\BCBL {}\ \BBA {} Aslam%
}{%
Kanoulas%
\ \protect \BOthers {.}}{%
{\protect \APACyear {2010}}%
}]{%
Kanoulas2010}
\APACinsertmetastar {%
Kanoulas2010}%
\begin{APACrefauthors}%
Kanoulas, E.%
, Dai, K.%
, Pavlu, V.%
\BCBL {}\ \BBA {} Aslam, J\BPBI A.%
\end{APACrefauthors}%
\unskip\
\newblock
\APACrefYearMonthDay{2010}{}{}.
\newblock
{\BBOQ}\APACrefatitle {Score Distribution Models: Assumptions, Intuition, and
  Robustness to Score Manipulation} {Score distribution models: Assumptions,
  intuition, and robustness to score manipulation}.{\BBCQ}
\newblock
\BIn{} \APACrefbtitle {Proceedings of {SIGIR} 2010} {Proceedings of {SIGIR}
  2010}\ (\BPGS\ 242--249).
\newblock
\APACaddressPublisher{New York, NY, USA}{ACM}.
\PrintBackRefs{\CurrentBib}

\bibitem [\protect \citeauthoryear {%
Kanoulas%
, Pavlu%
\BCBL {}\ \BBA {} Dai%
}{%
Kanoulas%
\ \protect \BOthers {.}}{%
{\protect \APACyear {2009}}%
}]{%
Kanoulas2009}
\APACinsertmetastar {%
Kanoulas2009}%
\begin{APACrefauthors}%
Kanoulas, E.%
, Pavlu, V.%
\BCBL {}\ \BBA {} Dai, K.%
\end{APACrefauthors}%
\unskip\
\newblock
\APACrefYearMonthDay{2009}{}{}.
\newblock
{\BBOQ}\APACrefatitle {{Modeling the score distributions of relevant and
  non-relevant documents}} {{Modeling the score distributions of relevant and
  non-relevant documents}}.{\BBCQ}
\newblock
\BIn{} \APACrefbtitle {Proceedings {ICTIR} 2009} {Proceedings {ICTIR} 2009}\
  (\BPGS\ 152--163).
\PrintBackRefs{\CurrentBib}

\bibitem [\protect \citeauthoryear {%
Kempthorne%
\ \BBA {} Doerfler%
}{%
Kempthorne%
\ \BBA {} Doerfler%
}{%
{\protect \APACyear {1969}}%
}]{%
Kempthorne}
\APACinsertmetastar {%
Kempthorne}%
\begin{APACrefauthors}%
Kempthorne, O.%
\BCBT {}\ \BBA {} Doerfler, T.%
\end{APACrefauthors}%
\unskip\
\newblock
\APACrefYearMonthDay{1969}{{\APACmonth{02}}}{}.
\newblock
{\BBOQ}\APACrefatitle {The behaviour of some significance tests under
  experimental randomization} {The behaviour of some significance tests under
  experimental randomization}.{\BBCQ}
\newblock
\APACjournalVolNumPages{Biometrika}{56}{2}{231--248}.
\PrintBackRefs{\CurrentBib}

\bibitem [\protect \citeauthoryear {%
Losada%
, Parapar%
\BCBL {}\ \BBA {} Barreiro%
}{%
Losada%
\ \protect \BOthers {.}}{%
{\protect \APACyear {2018}}%
}]{%
losadaetal17b}
\APACinsertmetastar {%
losadaetal17b}%
\begin{APACrefauthors}%
Losada, D.%
, Parapar, J.%
\BCBL {}\ \BBA {} Barreiro, A.%
\end{APACrefauthors}%
\unskip\
\newblock
\APACrefYearMonthDay{2018}{}{}.
\newblock
{\BBOQ}\APACrefatitle {A Rank Fusion Approach based on Score Distributions for
  Prioritizing Relevance Assessments in Information Retrieval Evaluation} {A
  rank fusion approach based on score distributions for prioritizing relevance
  assessments in information retrieval evaluation}.{\BBCQ}
\newblock
\APACjournalVolNumPages{Information Fusion}{}{39}{56--71}.
\PrintBackRefs{\CurrentBib}

\bibitem [\protect \citeauthoryear {%
Manmatha%
, Rath%
\BCBL {}\ \BBA {} Feng%
}{%
Manmatha%
\ \protect \BOthers {.}}{%
{\protect \APACyear {2001}}%
}]{%
Manmatha2001}
\APACinsertmetastar {%
Manmatha2001}%
\begin{APACrefauthors}%
Manmatha, R.%
, Rath, T.%
\BCBL {}\ \BBA {} Feng, F.%
\end{APACrefauthors}%
\unskip\
\newblock
\APACrefYearMonthDay{2001}{}{}.
\newblock
{\BBOQ}\APACrefatitle {{Modeling score distributions for combining the outputs
  of search engines}} {{Modeling score distributions for combining the outputs
  of search engines}}.{\BBCQ}
\newblock
\BIn{} \APACrefbtitle {Proceedings {SIGIR} 2001} {Proceedings {SIGIR} 2001}\
  (\BPGS\ 267--275).
\newblock
\APACaddressPublisher{New York, USA}{ACM Press}.
\PrintBackRefs{\CurrentBib}

\bibitem [\protect \citeauthoryear {%
Miettunen%
\ \BBA {} Nieminen%
}{%
Miettunen%
\ \BBA {} Nieminen%
}{%
{\protect \APACyear {2003}}%
}]{%
Miettunen2003}
\APACinsertmetastar {%
Miettunen2003}%
\begin{APACrefauthors}%
Miettunen, J.%
\BCBT {}\ \BBA {} Nieminen, P.%
\end{APACrefauthors}%
\unskip\
\newblock
\APACrefYearMonthDay{2003}{}{}.
\newblock
{\BBOQ}\APACrefatitle {The effect of statistical methods and study reporting
  characteristics on the number of citations: A study of four general
  psychiatric journals} {The effect of statistical methods and study reporting
  characteristics on the number of citations: A study of four general
  psychiatric journals}.{\BBCQ}
\newblock
\APACjournalVolNumPages{Scientometrics}{57}{3}{377--388}.
\PrintBackRefs{\CurrentBib}

\bibitem [\protect \citeauthoryear {%
Nelder%
\ \BBA {} Mead%
}{%
Nelder%
\ \BBA {} Mead%
}{%
{\protect \APACyear {1965}}%
}]{%
Nelder1965}
\APACinsertmetastar {%
Nelder1965}%
\begin{APACrefauthors}%
Nelder, J\BPBI A.%
\BCBT {}\ \BBA {} Mead, R.%
\end{APACrefauthors}%
\unskip\
\newblock
\APACrefYearMonthDay{1965}{{\APACmonth{01}}}{}.
\newblock
{\BBOQ}\APACrefatitle {{A Simplex Method for Function Minimization}} {{A
  Simplex Method for Function Minimization}}.{\BBCQ}
\newblock
\APACjournalVolNumPages{The Computer Journal}{7}{4}{308--313}.
\PrintBackRefs{\CurrentBib}

\bibitem [\protect \citeauthoryear {%
Parapar%
, Presedo-Quindimil%
\BCBL {}\ \BBA {} Barreiro%
}{%
Parapar%
\ \protect \BOthers {.}}{%
{\protect \APACyear {2014}}%
}]{%
Parapar2014}
\APACinsertmetastar {%
Parapar2014}%
\begin{APACrefauthors}%
Parapar, J.%
, Presedo-Quindimil, M\BPBI A.%
\BCBL {}\ \BBA {} Barreiro, A.%
\end{APACrefauthors}%
\unskip\
\newblock
\APACrefYearMonthDay{2014}{}{}.
\newblock
{\BBOQ}\APACrefatitle {Score distributions for Pseudo Relevance Feedback}
  {Score distributions for pseudo relevance feedback}.{\BBCQ}
\newblock
\APACjournalVolNumPages{Information Sciences}{273}{}{171 - 181}.
\PrintBackRefs{\CurrentBib}

\bibitem [\protect \citeauthoryear {%
Robertson%
}{%
Robertson%
}{%
{\protect \APACyear {2007}}%
}]{%
Robertson2007}
\APACinsertmetastar {%
Robertson2007}%
\begin{APACrefauthors}%
Robertson, S.%
\end{APACrefauthors}%
\unskip\
\newblock
\APACrefYearMonthDay{2007}{}{}.
\newblock
{\BBOQ}\APACrefatitle {On Score Distributions and Relevance} {On score
  distributions and relevance}.{\BBCQ}
\newblock
\BIn{} \APACrefbtitle {Proceedings {ECIR} 2007} {Proceedings {ECIR} 2007}\
  (\BPGS\ 40--51).
\newblock
\APACaddressPublisher{}{Springer Berlin Heidelberg}.
\PrintBackRefs{\CurrentBib}

\bibitem [\protect \citeauthoryear {%
Robertson%
, Kanoulas%
\BCBL {}\ \BBA {} Yilmaz%
}{%
Robertson%
\ \protect \BOthers {.}}{%
{\protect \APACyear {2013}}%
}]{%
Robertson:2013:MSD:2499178.2499181}
\APACinsertmetastar {%
Robertson:2013:MSD:2499178.2499181}%
\begin{APACrefauthors}%
Robertson, S.%
, Kanoulas, E.%
\BCBL {}\ \BBA {} Yilmaz, E.%
\end{APACrefauthors}%
\unskip\
\newblock
\APACrefYearMonthDay{2013}{}{}.
\newblock
{\BBOQ}\APACrefatitle {Modelling Score Distributions Without Actual Scores}
  {Modelling score distributions without actual scores}.{\BBCQ}
\newblock
\BIn{} \APACrefbtitle {Proceedings {ICTIR} 2013} {Proceedings {ICTIR} 2013}\
  (\BPGS\ 20:85--20:92).
\newblock
\APACaddressPublisher{New York, NY, USA}{ACM}.
\PrintBackRefs{\CurrentBib}

\bibitem [\protect \citeauthoryear {%
Sakai%
}{%
Sakai%
}{%
{\protect \APACyear {2016}}%
}]{%
Sakai:2016:TST:2911451.2914684}
\APACinsertmetastar {%
Sakai:2016:TST:2911451.2914684}%
\begin{APACrefauthors}%
Sakai, T.%
\end{APACrefauthors}%
\unskip\
\newblock
\APACrefYearMonthDay{2016}{}{}.
\newblock
{\BBOQ}\APACrefatitle {Two Sample T-tests for {IR} Evaluation: Student or
  Welch?} {Two sample t-tests for {IR} evaluation: Student or welch?}{\BBCQ}
\newblock
\BIn{} \APACrefbtitle {Proceedings {SIGIR} 2016} {Proceedings {SIGIR} 2016}\
  (\BPGS\ 1045--1048).
\newblock
\APACaddressPublisher{New York, USA}{ACM}.
\PrintBackRefs{\CurrentBib}

\bibitem [\protect \citeauthoryear {%
Sanderson%
\ \BBA {} Zobel%
}{%
Sanderson%
\ \BBA {} Zobel%
}{%
{\protect \APACyear {2005}}%
}]{%
Sanderson:2005}
\APACinsertmetastar {%
Sanderson:2005}%
\begin{APACrefauthors}%
Sanderson, M.%
\BCBT {}\ \BBA {} Zobel, J.%
\end{APACrefauthors}%
\unskip\
\newblock
\APACrefYearMonthDay{2005}{}{}.
\newblock
{\BBOQ}\APACrefatitle {Information Retrieval System Evaluation: Effort,
  Sensitivity, and Reliability} {Information retrieval system evaluation:
  Effort, sensitivity, and reliability}.{\BBCQ}
\newblock
\BIn{} \APACrefbtitle {Proceedings {SIGIR} 2005} {Proceedings {SIGIR} 2005}\
  (\BPGS\ 162--169).
\newblock
\APACaddressPublisher{New York, NY, USA}{ACM}.
\PrintBackRefs{\CurrentBib}

\bibitem [\protect \citeauthoryear {%
Savoy%
}{%
Savoy%
}{%
{\protect \APACyear {1997}}%
}]{%
Savoy1997495}
\APACinsertmetastar {%
Savoy1997495}%
\begin{APACrefauthors}%
Savoy, J.%
\end{APACrefauthors}%
\unskip\
\newblock
\APACrefYearMonthDay{1997}{}{}.
\newblock
{\BBOQ}\APACrefatitle {Statistical inference in retrieval effectiveness
  evaluation} {Statistical inference in retrieval effectiveness
  evaluation}.{\BBCQ}
\newblock
\APACjournalVolNumPages{Information Processing \& Management}{33}{4}{495 -
  512}.
\PrintBackRefs{\CurrentBib}

\bibitem [\protect \citeauthoryear {%
Smucker%
, Allan%
\BCBL {}\ \BBA {} Carterette%
}{%
Smucker%
\ \protect \BOthers {.}}{%
{\protect \APACyear {2007}}%
}]{%
Smucker:2007}
\APACinsertmetastar {%
Smucker:2007}%
\begin{APACrefauthors}%
Smucker, M\BPBI D.%
, Allan, J.%
\BCBL {}\ \BBA {} Carterette, B.%
\end{APACrefauthors}%
\unskip\
\newblock
\APACrefYearMonthDay{2007}{}{}.
\newblock
{\BBOQ}\APACrefatitle {A Comparison of Statistical Significance Tests for
  Information Retrieval Evaluation} {A comparison of statistical significance
  tests for information retrieval evaluation}.{\BBCQ}
\newblock
\BIn{} \APACrefbtitle {Proceedings {CIKM} 2007} {Proceedings {CIKM} 2007}\
  (\BPGS\ 623--632).
\newblock
\APACaddressPublisher{New York, NY, USA}{ACM}.
\PrintBackRefs{\CurrentBib}

\bibitem [\protect \citeauthoryear {%
Swets%
}{%
Swets%
}{%
{\protect \APACyear {1963}}%
}]{%
J.A.Sweets1963}
\APACinsertmetastar {%
J.A.Sweets1963}%
\begin{APACrefauthors}%
Swets, J\BPBI A.%
\end{APACrefauthors}%
\unskip\
\newblock
\APACrefYearMonthDay{1963}{}{}.
\newblock
{\BBOQ}\APACrefatitle {{Information Retrieval Systems}} {{Information Retrieval
  Systems}}.{\BBCQ}
\newblock
\APACjournalVolNumPages{Science}{141}{3577}{245--250}.
\PrintBackRefs{\CurrentBib}

\bibitem [\protect \citeauthoryear {%
Swets%
}{%
Swets%
}{%
{\protect \APACyear {1969}}%
}]{%
Swets1969}
\APACinsertmetastar {%
Swets1969}%
\begin{APACrefauthors}%
Swets, J\BPBI A.%
\end{APACrefauthors}%
\unskip\
\newblock
\APACrefYearMonthDay{1969}{}{}.
\newblock
{\BBOQ}\APACrefatitle {{Effectiveness of Information Retrieval Methods}}
  {{Effectiveness of Information Retrieval Methods}}.{\BBCQ}
\newblock
\APACjournalVolNumPages{American Documentation}{20}{}{72--89}.
\PrintBackRefs{\CurrentBib}

\bibitem [\protect \citeauthoryear {%
Urbano%
}{%
Urbano%
}{%
{\protect \APACyear {2016}}%
}]{%
Urbano2016}
\APACinsertmetastar {%
Urbano2016}%
\begin{APACrefauthors}%
Urbano, J.%
\end{APACrefauthors}%
\unskip\
\newblock
\APACrefYearMonthDay{2016}{Jun}{01}.
\newblock
{\BBOQ}\APACrefatitle {Test collection reliability: a study of bias and
  robustness to statistical assumptions via stochastic simulation} {Test
  collection reliability: a study of bias and robustness to statistical
  assumptions via stochastic simulation}.{\BBCQ}
\newblock
\APACjournalVolNumPages{Information Retrieval Journal}{19}{3}{313--350}.
\PrintBackRefs{\CurrentBib}

\bibitem [\protect \citeauthoryear {%
Urbano%
, Marrero%
\BCBL {}\ \BBA {} Mart\'{\i}n%
}{%
Urbano%
\ \protect \BOthers {.}}{%
{\protect \APACyear {2013}}%
}]{%
Urbano:2013}
\APACinsertmetastar {%
Urbano:2013}%
\begin{APACrefauthors}%
Urbano, J.%
, Marrero, M.%
\BCBL {}\ \BBA {} Mart\'{\i}n, D.%
\end{APACrefauthors}%
\unskip\
\newblock
\APACrefYearMonthDay{2013}{}{}.
\newblock
{\BBOQ}\APACrefatitle {A Comparison of the Optimality of Statistical
  Significance Tests for Information Retrieval Evaluation} {A comparison of the
  optimality of statistical significance tests for information retrieval
  evaluation}.{\BBCQ}
\newblock
\BIn{} \APACrefbtitle {Proceedings {SIGIR} 2013} {Proceedings {SIGIR} 2013}\
  (\BPGS\ 925--928).
\newblock
\APACaddressPublisher{New York}{ACM}.
\PrintBackRefs{\CurrentBib}

\bibitem [\protect \citeauthoryear {%
Van~Rijsbergen%
}{%
Van~Rijsbergen%
}{%
{\protect \APACyear {1979}}%
}]{%
Rijsbergen:1979}
\APACinsertmetastar {%
Rijsbergen:1979}%
\begin{APACrefauthors}%
Van~Rijsbergen, C\BPBI J.%
\end{APACrefauthors}%
\unskip\
\newblock
\APACrefYear{1979}.
\newblock
\APACrefbtitle {Information Retrieval} {Information retrieval}\
  (\PrintOrdinal{2nd}\ \BEd).
\newblock
\APACaddressPublisher{Newton, MA, USA}{Butterworth-Heinemann}.
\PrintBackRefs{\CurrentBib}

\bibitem [\protect \citeauthoryear {%
Voorhees%
\ \BBA {} Buckley%
}{%
Voorhees%
\ \BBA {} Buckley%
}{%
{\protect \APACyear {2002}}%
}]{%
Voorhees2002}
\APACinsertmetastar {%
Voorhees2002}%
\begin{APACrefauthors}%
Voorhees, E\BPBI M.%
\BCBT {}\ \BBA {} Buckley, C.%
\end{APACrefauthors}%
\unskip\
\newblock
\APACrefYearMonthDay{2002}{}{}.
\newblock
{\BBOQ}\APACrefatitle {{The Effect of Topic Set Size on Retrieval Experiment
  Error}} {{The Effect of Topic Set Size on Retrieval Experiment
  Error}}.{\BBCQ}
\newblock
\BIn{} \APACrefbtitle {Proceedings of the 25th Annual International {ACM SIGIR}
  Conference on Research and Development in Information Retrieval} {Proceedings
  of the 25th annual international {ACM SIGIR} conference on research and
  development in information retrieval}\ (\BPGS\ 316--323).
\PrintBackRefs{\CurrentBib}

\bibitem [\protect \citeauthoryear {%
Voorhees%
\ \BBA {} Harman%
}{%
Voorhees%
\ \BBA {} Harman%
}{%
{\protect \APACyear {2005}}%
}]{%
Voorhees:2005:TEE:1121636}
\APACinsertmetastar {%
Voorhees:2005:TEE:1121636}%
\begin{APACrefauthors}%
Voorhees, E\BPBI M.%
\BCBT {}\ \BBA {} Harman, D\BPBI K.%
\end{APACrefauthors}%
\unskip\
\newblock
\APACrefYear{2005}.
\newblock
\APACrefbtitle {{TREC}: Experiment and Evaluation in Information Retrieval}
  {{TREC}: Experiment and evaluation in information retrieval}.
\newblock
\APACaddressPublisher{}{The MIT Press}.
\PrintBackRefs{\CurrentBib}

\bibitem [\protect \citeauthoryear {%
Webber%
, Moffat%
\BCBL {}\ \BBA {} Zobel%
}{%
Webber%
\ \protect \BOthers {.}}{%
{\protect \APACyear {2008}}%
}]{%
Webber:2008:SPR:1458082.1458158}
\APACinsertmetastar {%
Webber:2008:SPR:1458082.1458158}%
\begin{APACrefauthors}%
Webber, W.%
, Moffat, A.%
\BCBL {}\ \BBA {} Zobel, J.%
\end{APACrefauthors}%
\unskip\
\newblock
\APACrefYearMonthDay{2008}{}{}.
\newblock
{\BBOQ}\APACrefatitle {Statistical Power in Retrieval Experimentation}
  {Statistical power in retrieval experimentation}.{\BBCQ}
\newblock
\BIn{} \APACrefbtitle {Proceedings of the 17th {ACM} Conference on Information
  and Knowledge Management} {Proceedings of the 17th {ACM} conference on
  information and knowledge management}\ (\BPGS\ 571--580).
\newblock
\APACaddressPublisher{New York, NY, USA}{ACM}.
\PrintBackRefs{\CurrentBib}

\bibitem [\protect \citeauthoryear {%
Zobel%
}{%
Zobel%
}{%
{\protect \APACyear {1998}}%
}]{%
Zobel:1998}
\APACinsertmetastar {%
Zobel:1998}%
\begin{APACrefauthors}%
Zobel, J.%
\end{APACrefauthors}%
\unskip\
\newblock
\APACrefYearMonthDay{1998}{}{}.
\newblock
{\BBOQ}\APACrefatitle {How Reliable Are the Results of Large-scale Information
  Retrieval Experiments?} {How reliable are the results of large-scale
  information retrieval experiments?}{\BBCQ}
\newblock
\BIn{} \APACrefbtitle {Proceedings {SIGIR} 1998} {Proceedings {SIGIR} 1998}\
  (\BPGS\ 307--314).
\newblock
\APACaddressPublisher{New York, NY, USA}{ACM}.
\PrintBackRefs{\CurrentBib}

\end{thebibliography}

\section*{Appendix}

The following pseudo-code computes the probability of a type I error, $p(Reject$ $H0|H0$ $is$ $true)$:

\begin{algorithm}[H]
		\scriptsize 
	\KwIn{A set of TREC systems \{$\mathcal{S}_1$,$\mathcal{S}_2$, \ldots, $\mathcal{S}_m$\}, a set of queries  \{$q_1$,$q_2$, \ldots,  $q_n$\},  a set of set of relevance judgements  \{$\mathcal{J}_1$,$\mathcal{J}_2$  \ldots, , $\mathcal{J}_n$\}, for each system $\mathcal{S}_j$ and query $q_i$ a set of scores $S_{j,i}=\{s_{j,i}^1, s_{j,i}^2, \ldots, s_{j,i}^{1000}\}$ and a significance level $\alpha$}	
	\For{$j \gets 1$ \textbf{to} $m$} {
		\For{$i \gets 1$ \textbf{to} $n$} {
			$mixture[j,i] \gets learnLogNormalMixture(S_{j,i},\mathcal{J}_i)$
		}
	}
	\For{$j \gets 1$ \textbf{to} $m$} {
		$rejectionWilcoxon \gets  0$\;
		$rejectionSign \gets  0$\;
		$rejectionTtest \gets  0$\;		
		$rejectionPermutation \gets  0$\;				
		$rejectionBootstrap \gets  0$\;				
		\For{$k \gets 1$ \textbf{to} $1000$} {
			\For{$i \gets 1$ \textbf{to} $n$} {	
				$S^1_{j,i} \gets sort(randomSample(mixture[j,i], 1000))$\;
				$S^2_{j,i} \gets sort(randomSample(mixture[j,i], 1000))$\;
				$ap^1[j,i] \gets ap(S^1_{j,i},mixture[j,i])$\;
				$ap^2[j,i] \gets ap(S^2_{j,i},mixture[j,i])$\;
			}	
			$rejectionWilcoxon \gets rejectionWilcoxon + testWilcoxon(ap^1[j],ap^2[j],\alpha)$\;
			$rejectionSign \gets rejectionSign + testSign(ap^1[j],ap^2[j],\alpha)$\;
			$rejectionTtest \gets rejectionTtest + testTtest(ap^1[j],ap^2[j],\alpha)$\;
			$rejectionPermutation \gets rejectionPermutation + testPermutation(ap^1[j],ap^2[j],\alpha)$\;
			$rejectionBootstrap \gets rejectionBootstrap + testBootstrap(ap^1[j],ap^2[j],\alpha)$\;			
		}
		$pRejectH0GivenH0[wilcoxon] \gets pRejectH0GivenH0[wilcoxon]  + (rejectionWilcoxon / 1000)$\;
		$pRejectH0GivenH0[sign] \gets pRejectH0GivenH0[sign]  + (rejectionSign / 1000)$\;
		$pRejectH0GivenH0[ttest] \gets pRejectH0GivenH0[ttest] + ()rejectionTtest / 1000)$\;
		$pRejectH0GivenH0[permutation] \gets pRejectH0GivenH0[permutation] + (rejectionPermutation / 1000)$\;
		$pRejectH0GivenH0[bootstrap] \gets pRejectH0GivenH0[bootstrap] + (rejectionBootstrap / 1000)$\;
	}
	$pRejectH0GivenH0[wilcoxon] \gets pRejectH0GivenH0[wilcoxon] / m$\;
	$pRejectH0GivenH0[sign] \gets pRejectH0GivenH0[sign] / m$\;
	$pRejectH0GivenH0[ttest] \gets pRejectH0GivenH0[ttest] / m$\;
	$pRejectH0GivenH0[permutation] \gets pRejectH0GivenH0[permutation] / m $\;
	$pRejectH0GivenH0[bootstrap] \gets pRejectH0GivenH0[bootstrap] / m $\;
	\caption{Pseudo-code for computing the probability of a type I error	}
	\label{algo:exactness}
\end{algorithm}

\noindent where  $m$ is the number of systems, $n$ is the number of queries, the $sort()$ method sorts in descending order, $randomSample(mixture[j,i], x)$ generates an array of size $x$ with random samples from the mixture, and $testNameOfThe\-Test(ap^1[j],ap^2[j],\alpha)$ returns 1 when the test determines rejection of $H0$ given the provided $alpha$ and 0 otherwise. 

\newpage

The following pseudo-code produces the power figures:

\begin{algorithm}[H]
	\scriptsize 
	\KwIn{A set of TREC systems \{$\mathcal{S}_1$,$\mathcal{S}_2$, \ldots, $\mathcal{S}_m$\}, a set of queries  \{$q_1$,$q_2$, \ldots,  $q_n$\},  a set of set of relevance judgements  \{$\mathcal{J}_1$,$\mathcal{J}_2$  \ldots, , $\mathcal{J}_n$\}, for each system $\mathcal{S}_j$ and query $q_i$ a set of scores $S_{j,i}=\{s_{j,i}^1, s_{j,i}^2, \ldots, s_{j,i}^{1000}\}$ and a significance level $\alpha$}
	\For{$j \gets 1$ \textbf{to} $m$} {
		\For{$i \gets 1$ \textbf{to} $n$} {
			$mixture[j,i] \gets learnLogNormalMixture(S_{j,i},\mathcal{J}_i)$
		}
	}
	\For{$j \gets 1$ \textbf{to} $m$} {
		$rejectionWilcoxon \gets  0$\;
		$rejectionSign \gets  0$\;
		$rejectionTtest \gets  0$\;		
		$rejectionPermutation \gets  0$\;				
		$rejectionBootstrap \gets  0$\;	
		\For{$h \gets 0.0$ \textbf{to} $0.30$} {			
			\For{$k \gets 1$ \textbf{to} $1000$} {
				\For{$i \gets 1$ \textbf{to} $n$} {	
					$S^1_{j,i} \gets sort(randomSample(mixture[j,i], 1000))$\;
					$mixture'[j,i] \gets mixture[j,i]$\;
					$mixture'[j,i].\mu_1 \gets mixture[j,i].\mu_1 \times (1+ h) $\;
					$S^2_{j,i} \gets sort(randomSample(mixture'[j,i], 1000))$\;
					$ap^1[j,i] \gets ap(S^1_{j,i},mixture[j,i])$\;
					$ap^2[j,i] \gets ap(S^2_{j,i},mixture'[j,i])$\;
				}	
				$rejectionWilcoxon \gets rejectionWilcoxon + testWilcoxon(ap^1[j],ap^2[j],\alpha)$\;
				$rejectionSign \gets rejectionSign + testSign(ap^1[j],ap^2[j],\alpha)$\;
				$rejectionTtest \gets rejectionTtest + testTtest(ap^1[j],ap^2[j],\alpha)$\;
				$rejectionPermutation \gets rejectionPermutation + testPermutation(ap^1[j],ap^2[j],\alpha)$\;
				$rejectionBootstrap \gets rejectionBootstrap + testBootstrap(ap^1[j],ap^2[j],\alpha)$\;			
			}
			$pRejectH0[wilcoxon,h] \gets pRejectH0[wilcoxon,h]  + (rejectionWilcoxon / 1000)$\;
			$pRejectH0[sign,h] \gets pRejectH0[sign,h]  + (rejectionSign / 1000)$\;
			$pRejectH0[ttest,h] \gets pRejectH0[ttest,h] + ()rejectionTtest / 1000)$\;
			$pRejectH0[permutation,h] \gets pRejectH0[permutation,h] + (rejectionPermutation / 1000)$\;
			$pRejectH0[bootstrap,h] \gets pRejectH0[bootstrap,h] + (rejectionBootstrap / 1000)$\;
		}
	}
	\For{$h \gets 0.0$ \textbf{to} $0.15$} {			
		$pRejectH0[wilcoxon,h] \gets pRejectH0[wilcoxon,h] / m$\;
		$pRejectH0[sign,h] \gets pRejectH0[sign,h] / m$\;
		$pRejectH0[ttest,h] \gets pRejectH0[ttest,h] / m$\;
		$pRejectH0[permutation,h] \gets pRejectH0[permutation,h] / m $\;
		$pRejectH0[bootstrap,h] \gets pRejectH0[bootstrap,h] / m $\;
	}
	\caption{Pseudo-code for computing the power plots of the statistical tests for a TREC edition}
	\label{algo:power}
\end{algorithm}

\noindent where with $mixture'[j,i].\mu_1 \gets mixture[j,i].\mu_1 \times (1+ h)$ we are altering the mean of the distribution of relevant documents in the mixture by a $h$ percentage.

\end{document}